\documentclass[useAMS,usenatbib,asm]{mn2e}
\usepackage{graphicx}
\usepackage{fleqn,times,color}
\usepackage[T1]{fontenc}
\usepackage{aecompl}
\newcommand{\tal}{\it et al. \rm}
\newcommand{\ksp}{km\,s$^{-1}$\,kpc$^{-1}$}
\newcommand{\gr}{^{\circ}}
\newcommand{\kms}{km\,s$^{-1}$}
\title{The analysis of realistic Stellar Gaia mock catalogues. I. Red Clump Stars as tracers of the central bar}
\author[Romero-G\'omez \tal]{M. Romero-G\'omez$^1$, F. Figueras$^1$, T. Antoja$^2$, H. Abedi$^1$, L. Aguilar$^3$ \\  \\
$^1$Dept. d'Astronomia i Meteorologia, Institut de Ci\`encies del Cosmos, Universitat de Barcelona, IEEC, Mart\'{i} i Franqu\`es 1, E08028 Barcelona, Spain,\\ Email: mromero{@}am.ub.es\\
$^2$Research and Scientific Support Office, European Space Agency (ESA-ESTEC), PO Box 299, NL-2200 AG Noordwijk, the Netherlands\\
$^3$Instituto de Astronom\'{i}a, Universidad Nacional Aut\'onoma de Mexico, Apdo. 
Postal 877, Ensenada, 22800, Baja California, Mexico.\\
}

\date{Received }

\begin{document}

\maketitle

\begin{abstract}
In this first paper we simulate the population of disc Red Clump stars to 
be observed by Gaia. We generate a set of test particles and we 
evolve it in a 3D barred Milky Way like galactic potential. We assign 
physical properties of the Red Clump trace population and a 
realistic 3D interstellar extinction model. We add Gaia observational 
constraints and an error model according to the pre-commissioning scientific 
performance assessments. We present and analyse two mock catalogues, offered 
to the community, that are an excellent test bed for testing 
tools being developed for the future scientific exploitation of Gaia data. 
The first catalogue contains stars up to Gaia G$\sim$20, while the second is 
the subset containing Gaia radial velocity data with a maximum error of
$\sigma_{V_r}=10$\kms. 
Here we present first attempts to characterise the density structure of 
the Galactic bar in the Gaia space of observables. The Gaia large  
errors in parallax and the high interstellar extinction in the inner parts 
of the Galactic disc prevent us to model the bar overdensity. 
This result suggests the need to combine Gaia and IR data to 
undertake such studies. We find that IR photometric distances
for this Gaia sample allow us to recover the Galactic bar orientation angle 
with an accuracy of $\sim 5\gr$.
 
\end{abstract}

\begin{keywords}
Galaxy -- structure -- kinematics -- barred galaxies
\end{keywords}

\section{Introduction}
\label{sec:intro}
Special emphasis has been put recently on characterising the Galactic disc 
and its non-axisymmetric structures. Upcoming big surveys, such as Gaia (ESA), 
will provide us with the opportunity to increase the extent of the region with 
detailed kinematic information from few hundreds of parsecs to few 
kiloparsecs and with unprecedented precision. This larger volume of study 
makes it reasonable to question whether Gaia will be able to characterise 
the Galactic bar and the spiral arms, both in density, by determining 
overdensities, and in kinematics, by analysing the imprints that these 
structures leave in the velocity space 
\citep[e.g.][]{deh00,fux01,gar10,min10,ant11} .
One of the regions of great interest in the galactic disc is the end of the bar
region. It can provide information regarding the link between the spiral
arms and the bar.
Besides, this is a possible line-of-sight to be observed by the Gaia-ESO 
Survey.

In this paper we provide mock catalogues which are a suitable test bed for 
such studies. Several analysis have already started using these catalogues, 
for example the use of the vertex deviation to determine the resonant radii of 
galaxy discs \citep{roc14}, or determining the tilt and twist 
angles of the Galactic warp \citep{abe14}. Single population catalogues 
generated using test particle simulations together with a realistic extinction 
map, as proposed here, have demonstrated to be well suited to perform these 
kind of studies. The advantage of using test particles evolved in a realistic 
galactic potential, instead of samples generated from at present Galaxy models 
(e.g. the Besan\c{c}on Galaxy Model \citep{cze14}), is that the stars have 
been evolved according to a known galactic potential and have inherited the 
information on both density and kinematics, that is the stars are in 
statistical equilibrium with the potential imposed. With respect to N-body
simulations, the advantages are that we control the potential used and that
we can choose the parameters of the potential to resemble that of the 
Milky Way. Thus, by changing the parameters of the potential, such as disc 
mass, bar length, bar pattern speed, we obtain new mock catalogues fulfilling 
the new kinematic imprints of the mass model imposed. This fact allows us to 
consider the inverse problem once we have the Gaia data. That is, determining 
the free parameters of the potential whose characteristics match the observed 
data. 

Here we present two mock catalogues of Red Clump stars with 
the set of observables that Gaia will provide\footnote{Available upon request 
to the running author}. The first contains all such disc stars up to Gaia G 
magnitude of $20$, for which Gaia will provide astrometric and photometric 
data. The second catalogue is the subset containing also Gaia radial 
velocity data. This reduces the sample to stars up to magnitude G approximately 
$16$. The potential used to generate the catalogues 
is 3D, with a rotation curve that matches that of the Milky Way and it includes
the expected non-axisymmetric structure, i.e. the Galactic bar. During
the integration, response spiral arms arise. However, we will not discuss
them in this paper. The number of particles has been set to match the 
local surface density of the Red Clump population. 

It has been long established that the Red Clump K-giants (hereafter, RC) are 
a good tracer population suitable to study the Galactic structure. First, RC 
stars are abundant enough and sufficiently bright \citep{pac98,sta98,zas13}. 
Second, theoretical models predict that RC stars absolute luminosity barely 
depends on their age and chemical composition. RC stars have been used in 
various studies of the Galactic disc, for example 
to estimate the distance to the Galactic centre \citep{pac98}, to 
propose the presence of a Long bar \citep{lop07}, or for the analysis of 
streaming motions in the disc \citep{wil13}. Gaia works in the optical range 
of spectra, with the known limitation of not reaching deeply in the galactic 
disc. Furthermore, Gaia's spectrograph is not able to provide radial 
velocities for the faintest stars. Undergoing surveys in the infra-red, 
therefore, become necessary to complement Gaia data, e.g. APOGEE (Apache 
Point Observatory Galactic Evolution Experiment) \citep{eis11}, from which 
accurate radial velocities are measured, and precise distances to the RC 
stars are provided \citep{bov14}. The authors select the 
RC stars from the APOGEE survey based on their position in 
colour-metallicity-surface-gravity-effective-temperature 
space using a new method calibrated using stellar-evolution models and 
high-quality asteroseismology data. The particular narrow position of the 
RC stars in the colour-metallicity-luminosity space allows the authors to 
assign distances to the stars with a precision of $5$ to $10\%$, which
is better than the Gaia precision far from the Solar neighbourhood
(see Sect.~\ref{sec:par_error}). In this work, therefore, we also consider 
observed distances with IR errors, instead of parallax errors. 

By either using the astrometric, photometric and spectroscopic error models 
defined by the Gaia mission before commissioning or relative distance errors 
provided by photometric distances, we convolve our model stars into 
observables. The mock catalogues presented here can be easily readjusted to 
future error models after commissioning, or first and second data releases. 

As a first exploitation of these mock catalogues, we study whether Gaia
will be able to detect and characterise the Galactic bar, by either directly 
detecting its overdensity, or by analysing its imprint in the kinematic space
(Romero-G\'omez et al, in preparation). Regarding the bar overdensity, recent 
papers \citep{mar11,rom11} rise the possibility that only one long 
boxy-bulge bar is present in the Milky Way, contrary to what other studies 
suggest \citep{ham00,ben05,lop07}, namely a triaxial bulge plus a misaligned 
long bar. Will Gaia be able to determine the bar characteristics? Although we 
know that Gaia data will not reach the Bulge region, we will extend our study 
from the Solar Neighbourhood to a spherical region of about $4-5$kpc, to which 
we refer as the Gaia sphere. Can Gaia detect the dynamical effects of the 
bar(s) in the Gaia sphere? 

This paper is organized as follows. First, in Sect.~\ref{sec:simu} we describe 
our 3D test particle simulation, namely the initial conditions, the 
gravitational potential model, the integration process and the construction 
of the two mock catalogues. In Sect.~\ref{sec:RCmock},
we analyse the surface density, the distribution of the parallax accuracies
and the distribution of the radial and tangential velocities accuracies of 
the catalogues. Then, in Sect.~\ref{sec:internal} we present the 
characterisation of the Galactic bar in the space of observables and the
Gaia possibilities to detect it, suggesting a necessary link with 
IR surveys. Finally, our conclusions are presented in Sect.~\ref{sec:conc}.

\section{The simulations}
\label{sec:simu}
 Most of the previous studies using test particle simulations 
 with a bar potential were 2D and focused 
on the kinematics of regions near the Sun (e.g. 
\citealt{deh00,fux01,gar10,min10,ant11}).  Only recent papers, such 
as \citet{mon14} start using 3D test particle simulations with similar goals
as ours, this is, we want to model the signatures 
of the Galactic bar at large distances from the Sun. Our simulation aims to 
mimic the present spatial and kinematic distribution 
of the disc RC population in 3D and over a wide region of the Galaxy. 

The total galactic potential consists of an 
axisymmetric component plus a bar-like potential. We choose the \citet{all91} 
potential for the axisymmetric component, which consists of the superposition 
of a Miyamoto-Nagai disc, a spherical bulge and a spherical halo. The free 
parameters of the axisymmetric potential are chosen so that the rotation
curves matches that of the Milky Way \citep{all91}.

In Sect.~\ref{sec:ic}, we describe the initial conditions, while in 
Sect.~\ref{sec:model} we give the details of the 3D galactic barred model 
considered. In Sect.~\ref{sec:int}, we describe the integration process. 
Finally, in Sect.~\ref{sec:RCpop}, we assign the physical properties of
the RC population and Gaia errors.

\subsection{Setting up the initial conditions}
\label{sec:ic}
The initial conditions follow the density distribution of the Miyamoto-
Nagai disc \citep{miy75} with the parameters set in \citet{all91}. They  
are generated using the Hernquist method \citep{Hernquist.93}. 
The velocity field is approximated by gaussians, whose parameters are obtained
from the first order moments of the collisionless Boltzmann equation, 
simplified by the epicyclic approximation. The details are shown in the 
Appendix A of the paper.

We consider a single population with the characteristics of the RC 
K-giants. That is, we fix the velocity dispersions at the Sun position to be 
$\sigma_U=30.3$\kms, $\sigma_V=23.6$\kms and $\sigma_W=16.6$\kms
\citep[and references therein]{bin08}, and a constant scale-height 
of $300$pc \citep{rob86}. The asymmetric drift has been taken into 
account when computing the tangential component of the velocity. 

As previously mentioned, the number of particles in the disc has been 
set to match the surface density of RC stars in a local 
neighbourhood. We consider a cylinder for all $z$ of radius $100$pc 
centred in the Solar position and by using the new Besan\c{c}on Galaxy Model 
\citep{cze14}, we obtain a surface density of stars of $0.05614$stars/pc$^2$
(Czekaj, private communication). This surface density is imposed to the set
of test particlesobtained after relaxation 
(see Sect.~\ref{sec:int}). The suitable amount of disc RC stars that matches 
this surface density is $57\times 10^6$, and this is the number of particles
we will use in our simulations.

As in \citet{all91}, we assume the Sun is located at $R_{\odot}=8.5$kpc, 
the local standard of rest rotates with a circular velocity of 
$V_c(R_{\odot})=220$\kms and the Sun peculiar velocity is 
$(U,V,W)_{\odot}=(10.,5.25,7.17)$\kms \citep{deh98}. Throughout the paper,
the Galactic Centre is located in the origin of coordinates and the Sun is 
located on the negative x-axis.

\subsection{The bar model}
\label{sec:model}
We aim to model the Galactic bar as a bar with a boxy/bulge, i.e. the 
COBE/DIRBE triaxial bulge, plus the long bar. For this, we use a simple
model which consists of the superposition of two Ferrers ellipsoids 
\citep{fer77} with non-homogeneity index equal to $n=1$ \citep{rom11,mar11}.  
The main parameters of the models are fixed to values within observational 
ranges (see \citet{rom11}). For the COBE/DIRBE bulge we set the semi-major 
axis to $a=3.13$kpc and the axes ratios to $b/a=0.4$ and $c/a=0.29$. The mass 
is $M_{CB}=4.5\times\,10^{9}$M$_{\odot}$. The length of the Long bar 
is set to $a=4.5$kpc and the axes ratios to $b/a=0.15$ and $c/a=0.026$. 
The mass of the long bar is fixed to $M_{LB}=2.5\times\,10^{9}$
M$_{\odot}$. So finally we obtain a boxy/bulge type of bar with total mass 
equal to $M_b=7.\times10^{9}$M$_{\odot}$. Both major axes are aligned 
on the x-axis of the rotating reference system and we will refer to the 
superposition of these two ellipsoids as the Galactic bar. 

The Galactic bar is oriented at $20\gr$ from the Sun-Galactic Centre line. 
It rotates at a constant angular speed of $50$\ksp around the short z-axis. 
This value is within the range accepted for the COBE/DIRBE bar of the Milky 
Way \citep{ger11}.

\subsection{The integration process}
\label{sec:int}
We obtain the final mock catalogue following three steps. We first integrate 
the disc initial conditions in the axisymmetric \citet{all91} potential alone, 
to allow particles to reach a reasonable state of statistical equilibrium with
the total axisymmetric component. After some trial and error, and being
conservative, we opted for an integration time of $10\,Gyr$. 

Secondly, the non-axisymmetric 
component is introduced adiabatically in four bar rotations
($T_{grow}\sim 500\,Myr$). During this time, we require that the total mass 
of the system ($9\times 10^{11}\,M_{\odot}$) remains constant. To do 
this we consider a progressive mass transfer from the initial spherical bulge of
the Allen \& Santill\'an model to the bar in the following way:
$$\Phi_T(x,y,z,t)=\Phi_d+\Phi_h+(1-f(t)f_0)\Phi_{bul}+f(t)\Phi_b,$$
where $\Phi_T$ denotes the total potential. $\Phi_d$, $\Phi_h$ and 
$\Phi_{bul}$ are the disc, halo and bulge axisymmetric components of the 
\citet{all91} potential, respectively, and $\Phi_b$ denotes the bar potential. 
For the time function $f(t)$ we adopt the same fifth degree polynomial of time, 
$t$, as in Eq.(4) of \citet{deh00}. It has continuous derivatives 
guaranteeing a smooth transition from the non-barred to the barred state. 
$f_0$ is defined as $f_0=M_b/M_{bul}$. Therefore, when $t=T_{grow}$, 
we have a bar of $M_b=7.\times10^{9}$M$_{\odot}$ and a residual
axisymmetric bulge of mass $M_{bul}=7.\times10^{9}$M$_{\odot}$. Therefore, in 
the central part of the Galaxy there is a boxy/bulge type of bar plus an 
axisymmetric bulge with a total mass $1.4\times10^{10}$M$_{\odot}$. 

Finally, once the bar has been introduced, we integrate the particles in the
total potential for another four bar rotations, again to allow the particles
to reach a reasonable state of statistical equilibrium now with the final
potential. In the following sections, we will analyse, the RC population in 
this final snapshot. We name this whole particle set as RC-all.  We want to 
stress that the samples presented here are not a self-consistent dynamical
solution, but rather a tracer population that has reached, to a large
extent and by construction, statistical equilibrium with the final
assumed potential.

\begin{figure}
\begin{center}
\includegraphics[scale=0.25]{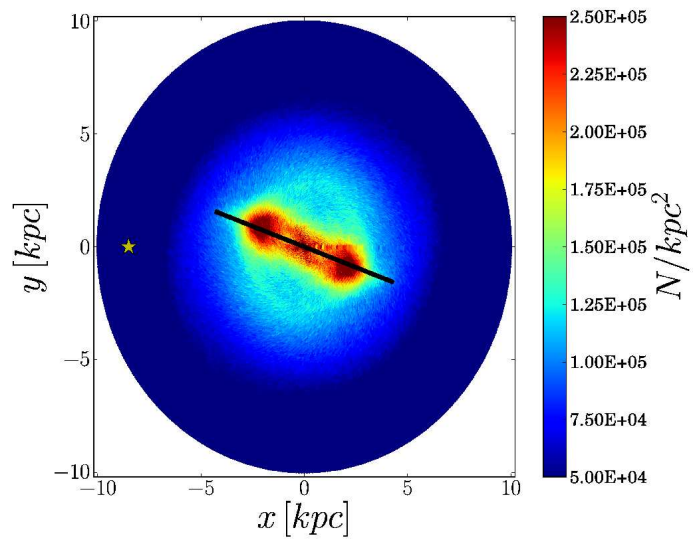}
\end{center}
\caption{Surface density (Number of stars / $kpc^2$) in each 
of the regions of the full sample model (RC-all). The yellow star shows the 
Sun position and the black solid line marks the position of the bar.}
\label{fig:Nvol_model}
\end{figure}

We show in Fig.~\ref{fig:Nvol_model} the surface density map 
corresponding to the RC-all sample. In this and following plots we divide 
the sample in galactocentric cylindrical bins $(R,\theta)$ of size 
$100$pc $\times 0.72\gr$, where $R\in [0,10]$kpc and 
$\theta \in [0,360\gr]\,$. In the case of Fig.~\ref{fig:Nvol_model}, 
we can clearly see the central non-axisymmetric component, i.e. the bar, 
in the surface density. 

\subsection{The RC physical parameters assignment}
\label{sec:RCpop}

The derivation of the Gaia G magnitude and the computation of the Gaia 
astrometric, photometric and spectrophotometric standard errors for the RC 
stars are calculated following the 
strategy outlined in the Gaia (ESA) webpage\footnote{\texttt{http://www.cosmos.esa.int/web/gaia/science-performance}}. This page is regularly updated according 
to the scientific performance assessments for Gaia during mission operation. 
The pre-launch performance models have been used in this paper. New 
prescriptions proposed after the commissioning phase (ended July 2014) have 
been recently published and more changes are expected before 
the first Gaia Data Release (mid 2016). These changes will be 
regularly applied to the catalogues presented here.

Our test particles are characterised as Red Clump K-giant stars, 
that is K0-1\,III stars. Here we 
assume they have an absolute magnitude  of $M_K=-1.61$ \citep{alv00} without 
intrinsic dispersion in brightness and intrinsic colors of $(V-I)_o=1.0$ and
$(V-K)_o=2.34$ \citep{alv00}. From the previous section, we have assigned a
galactocentric position to each star, which we can transform to heliocentric 
coordinates $(d,l,b)$, where $d$ is the heliocentric distance in kpc and
$(l,b)$ are the galactic longitude and latitude, respectively.
The 3D extinction model of \citet{dri03} using scaling factors has been used to
assign a visual absorption $A_V$ to the stars. The authors compute the 
scaling factors to correct the dust column density of the smooth model to 
account for small scale structure of the dust and gas not considered in the 
model. They are direction dependent factors based on the FIR residuals 
between the DIRBE 240$\mu m $ data and the predicted emission of the 
parametric dust distribution model. We assume then the extinction law 
from \citet{car89} ($A_K=0.114A_V$ and $A_I=0.479A_V$) to finally compute 
the apparent magnitude in V and the observed (V-I) colour index. 
The Gaia magnitude, G, and the $(B_p-R_p)$ colours are computed from a third 
degree polynomial fit depending on the apparent magnitude V and the (V-I) 
colour (see Table 3 of \citet{jor10}). 

The end-of-mission errors in astrometry depend on the magnitude G of the star 
and its observed (V-I) colour (see Gaia Science Performance webpage mentioned 
above). They also vary over the sky as a 
result of the scanning law due to the different number of transits at the end
of the mission (see Table 6 in the Gaia Science Performance webpage). Radial 
velocities will be obtained for stars brighter than $V\sim 17$ mag through 
Doppler-shift measurements by the Radial Velocity Spectrometer (RVS).

\section{The RC mock catalogues}
\label{sec:RCmock}
Once we have assigned the physical parameters of the RC stars to the 
simulated test particles, we obtain two simulated catalogues from our 
full disc simulation (RC-all). Both catalogues include the effects of the
absorption computed using the 3D Drimmel extinction model with scaling
factors mentioned above. 
 
The first catalogue, namely the RC-G20 sample, contains all stars with 
magnitude $G\le 20$, as this is the limiting magnitude for Gaia
astrometric and spectrophotometric data. The second catalogue, called RC-RVS, 
contains only stars having a radial velocity error $\sigma_{V_r}\le 10$\kms.
According to the Gaia Science Performance, the model for the radial velocity 
errors is valid for stars with $G_{rvs}\le 16.1$\footnote{The spectrometer 
operates in the region of the CaII triplet, that is 
$847-874\,\mu m$, and the integrated flux can be seen as measured with a 
photometric narrow band, $G_{rvs}$ magnitude.}, and the analytical expression 
for the error is an exponential function that depends on the magnitude V of 
the star and its spectral type. 
We add the Gaia errors to the astrometric and photometric variables and to
the radial velocities of both samples, obtaining the observed mock 
catalogues, RC-G20-O and RC-RVS-O, respectively, where O stands for 
observed. In Table~\ref{tab:Nstars}, we summarize the total number of stars 
of the samples considered. Cutting the RC-all sample to the stars that Gaia 
will see means reducing the total number of particles by a half. Furthermore, 
considering a sample with good radial 
velocities, RC-RVS, reduces the number of particles in the sample by a 
factor of 7.

\begin{figure}
\begin{center}
\includegraphics[scale=0.3]{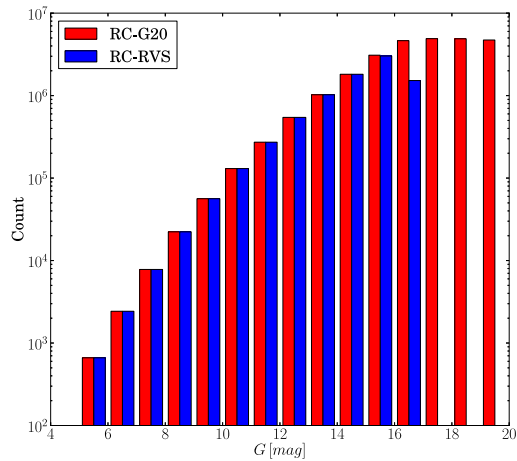}
\end{center}
\caption{Histograms in logscale of the magnitude $G$ for the RC-G20 (red) 
and RC-RVS (blue) samples.}
\label{fig:magG}
\end{figure}

Figure.~\ref{fig:magG} shows the histograms in the magnitude $G$ for the RC-G20 
sample (red) and for the RC-RVS sample (blue). The latter, which is obviously 
included in RC-G20, shows a limiting magnitude $G$ of about 16 mag, which is 
the result of the cut in radial velocity error of $\sigma_{V_r}\le 10$\kms. 
The flattening in the number of particles at faint magnitudes in the RC-G20 
sample is due to the combination of two facts: the effect of the spatial
density distribution and the high interstellar extinction at large distances 
in the plane.

\begin{table}
\begin{center}
\begin{tabular}{|c|c|}
\hline
\hline
 & Number of stars \\
\hline
\hline
RC-all & $57\times 10^6$ \\
\hline
RC-G20 & $26\times 10^6$ \\
\hline
RC-RVS & $8.5\times 10^6$ \\
\hline
\hline
\end{tabular}
\caption{Total number of stars of the samples used in the
paper, namely RC-all with all the particles in the disc, RC-G20 with
all the particles with Gaia G magnitude up to 20 mag and 
RC-RVS with all the particles with radial velocity error 
$\sigma_{V_r}\le 10$\kms. }
\label{tab:Nstars}
\end{center}
\end{table}

However, as discussed further in Sect.~\ref{sec:bar}, one 
of the advantages of using the RC population is that we can complement the 
Gaia data with other ongoing IR surveys, such as APOGEE or UKIDSS. As 
previously mentioned, distances derived from photometry will be more 
precise than trigonometric parallaxes for stars far from the Gaia sphere. 
\citet{bov14} derived spectro-photometric distances using high-resolution 
spectroscopic APOGEE data and NIR and mid-IR photometry. The authors estimated 
that distances for RC stars can be derived with an accuracy of $5$ to $10\%$. 
At present, the APOGEE survey covers only specific regions in the sky so this 
data would not be available for the full sky Gaia data. Photometric distances 
computed using only IR photometry will be needed \citep[e.g.][]{cab07}.
This second strategy suffers from contamination from non-RC stars, which 
would degrade distance estimation introducing some systematic trends. 
Recently, \citet{lop14} reported that for faint stars, the contamination  
due to non-RC stars could reach $20\%$. On the other hand, using 2MASS 
photometry and assuming an intrinsic dispersion of $\sigma_K=0.22$ for RC stars 
\citep{alv00}, by error propagation of the distance modulus, 
$M_K=m-5\log10 (d)+5-A_K$, a relative error in distance of about $10\%$ 
is estimated \citep{mon14}. Taking into account all these considerations, 
a relative error in distance of $10\%$ seems reasonable and it will be 
assumed here for our RC IR-distances. Therefore, in RC-G20-IR we convolve 
the RC-G20 mock catalogue with IR photometric distances.

\subsection{The RC surface density}
\label{sec:RCden}

\begin{figure*}
\begin{center}
\includegraphics[scale=0.22]{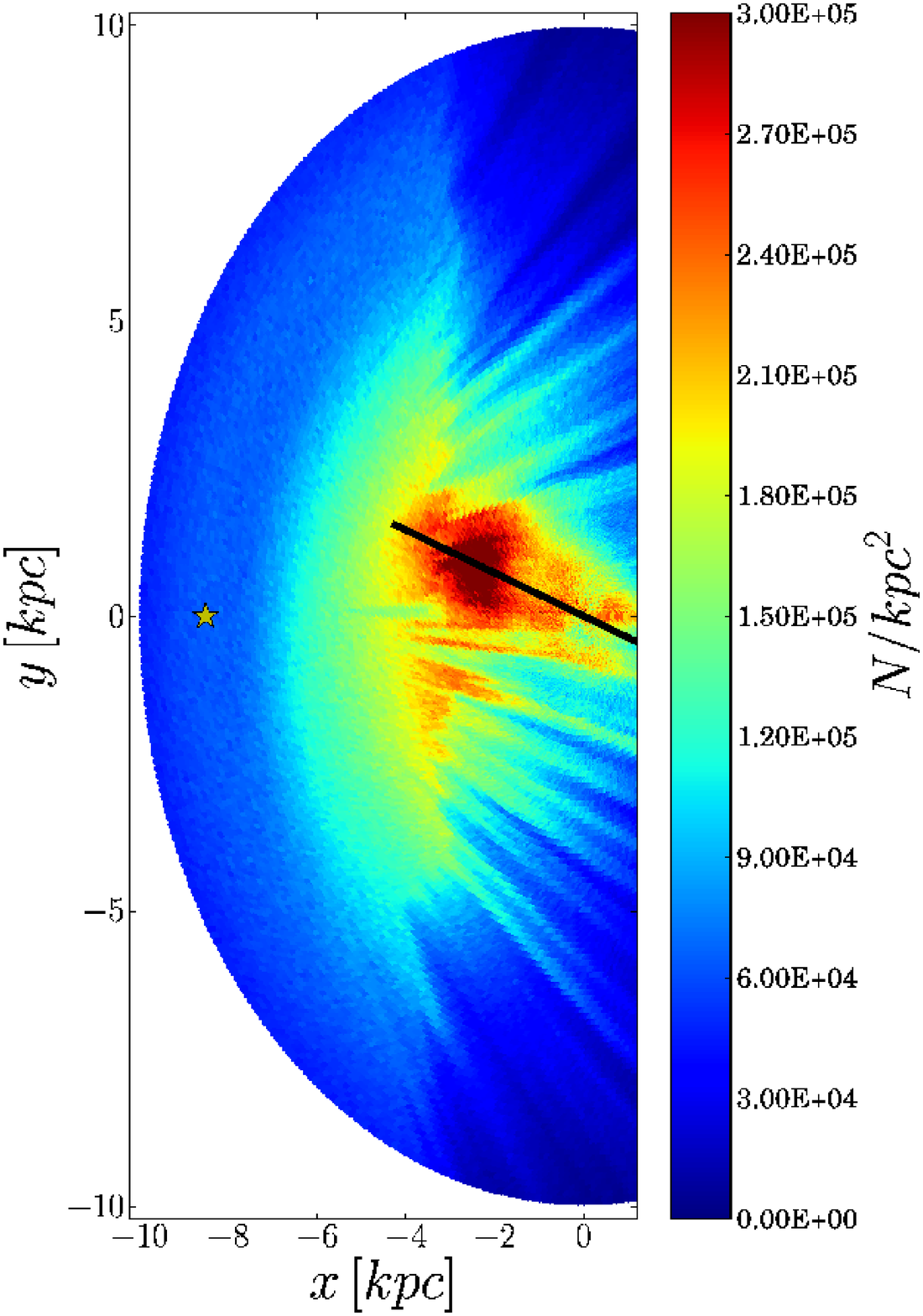}\hspace{.3cm}
\includegraphics[scale=0.22]{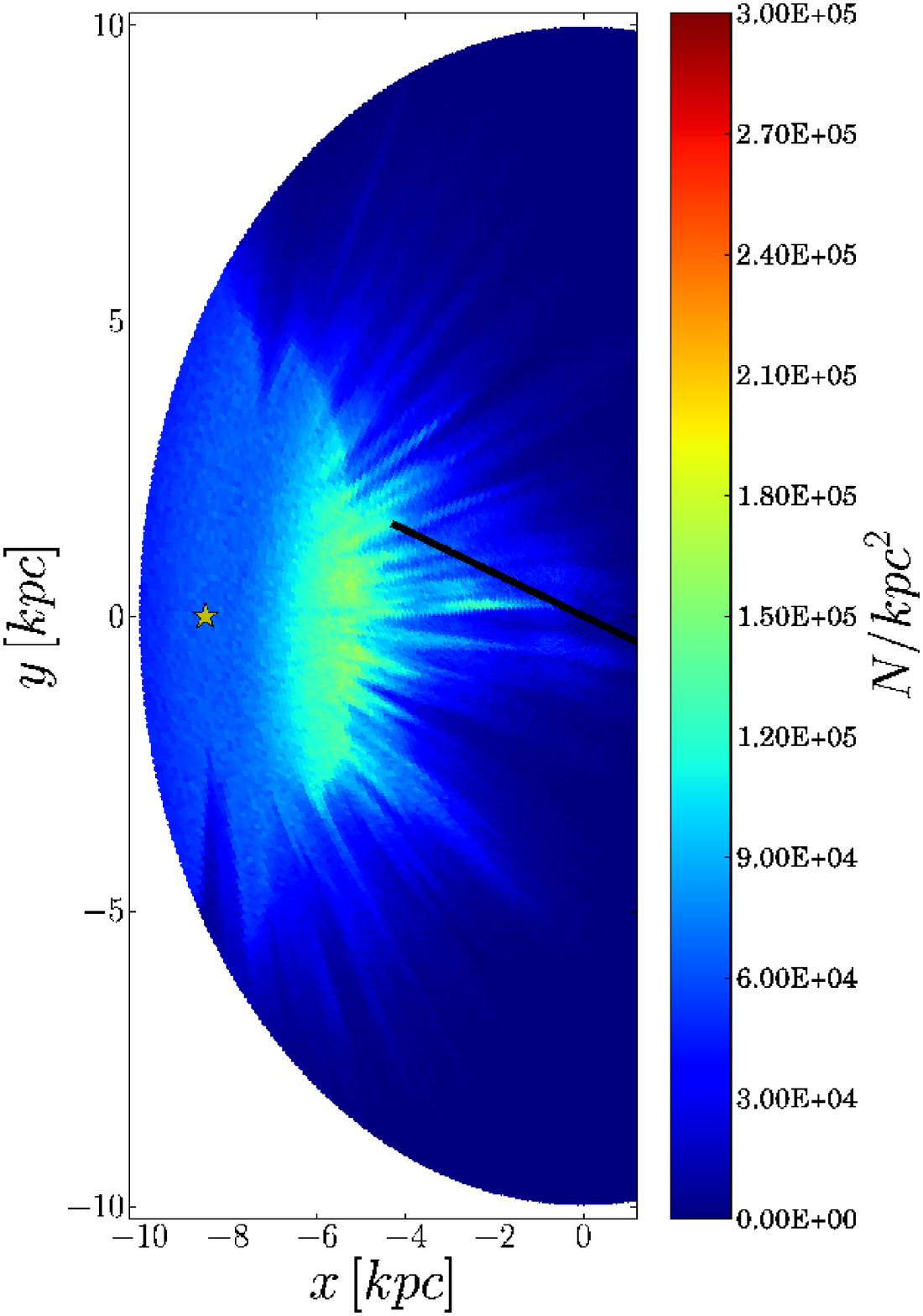}\\
\includegraphics[scale=0.22]{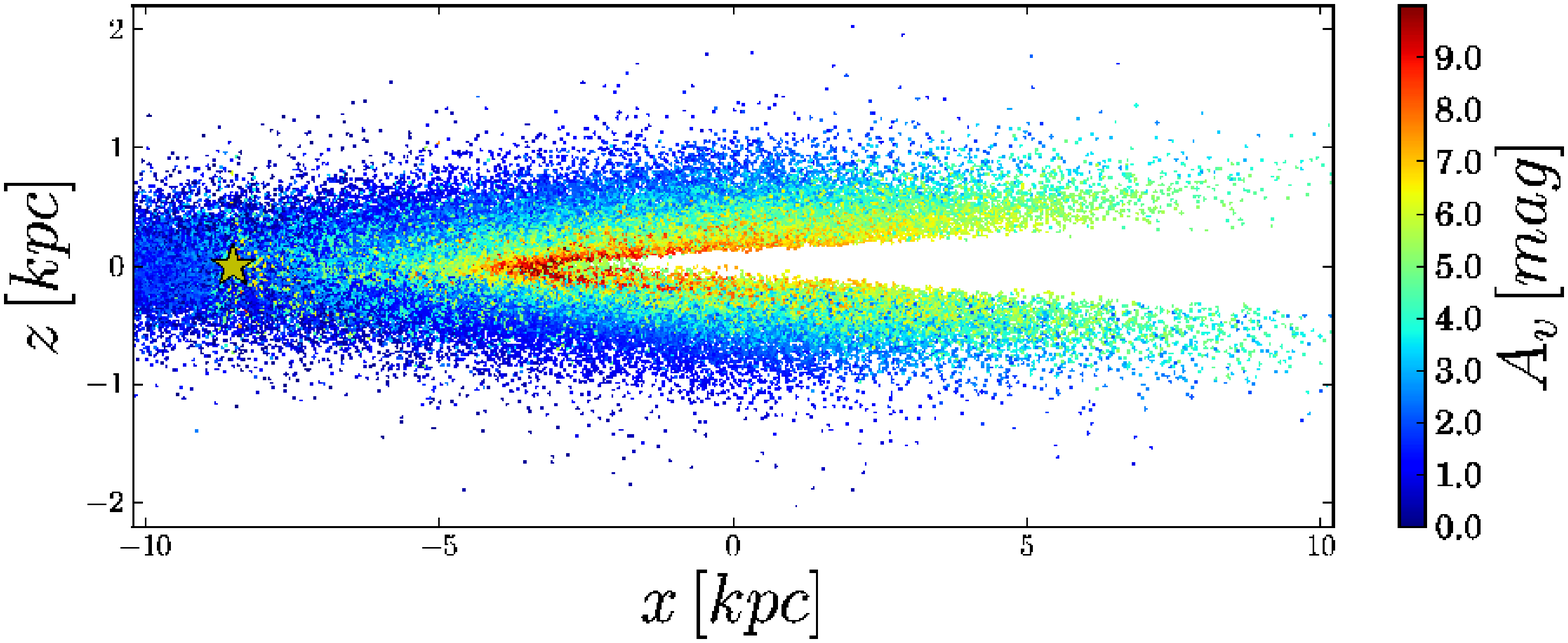}
\includegraphics[scale=0.22]{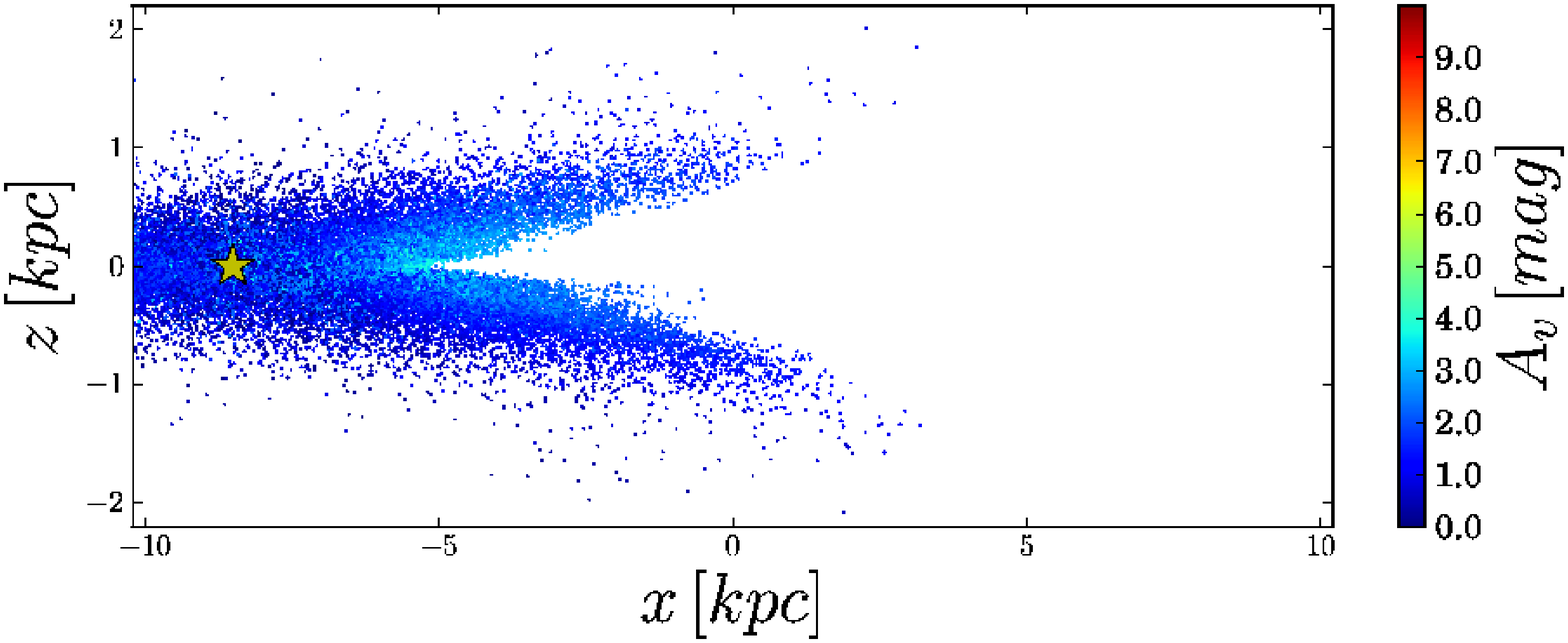}
\end{center}
\caption{The RC-G20 and RC-RVS samples. First row: As in 
Fig.~\ref{fig:Nvol_model} for the RC-G20 and
RC-RVS samples, respectively. Note that the stars are distributed using 
their real distances, i.e. not affected by errors. The yellow star in both 
panels shows the position of the Sun and the black solid line marks the 
position and length of the bar. Second row:
$(x,z)$ projection, the colour indicates the absorption in V given by the 
\citet{dri03} model.  }
\label{fig:Nvol_samples}
\end{figure*}

In Figure~\ref{fig:Nvol_samples} we show the surface density of the RC-G20 
and the RC-RVS samples. This gives us an estimation of the number 
of RC stars that we can expect in the Gaia catalogue in different locations in 
the disc. Here the stars are distributed using their real 
distances, i.e. not affected by errors.  The surface density looks 
different in each sample since both samples have a different magnitude cut. 
Taking into account all these facts, Gaia will be able to provide proper
motions for the RC stars located at the end of the bar with the RC-G20 sample 
but not radial velocities (left panel). 
However, as expected, the number of particles in this region decreases when we 
require the particles to have good radial velocities, that is for the RC-RVS 
sample (right panel).

In the bottom panels of Fig.~\ref{fig:Nvol_samples}, we show the distribution
of stars in $(x,z)$ plane coloured according to the absorption in V, 
 $A_V$. As expected, particles closer to the Galactic plane have higher 
absorption. In both cases, a $65\%$ of the particles in the sample lie within 
$|z|<300$pc. 

\subsection{Parallax precision distribution}
\label{sec:par_error}
\begin{figure}
\begin{center}
\includegraphics[scale=0.17]{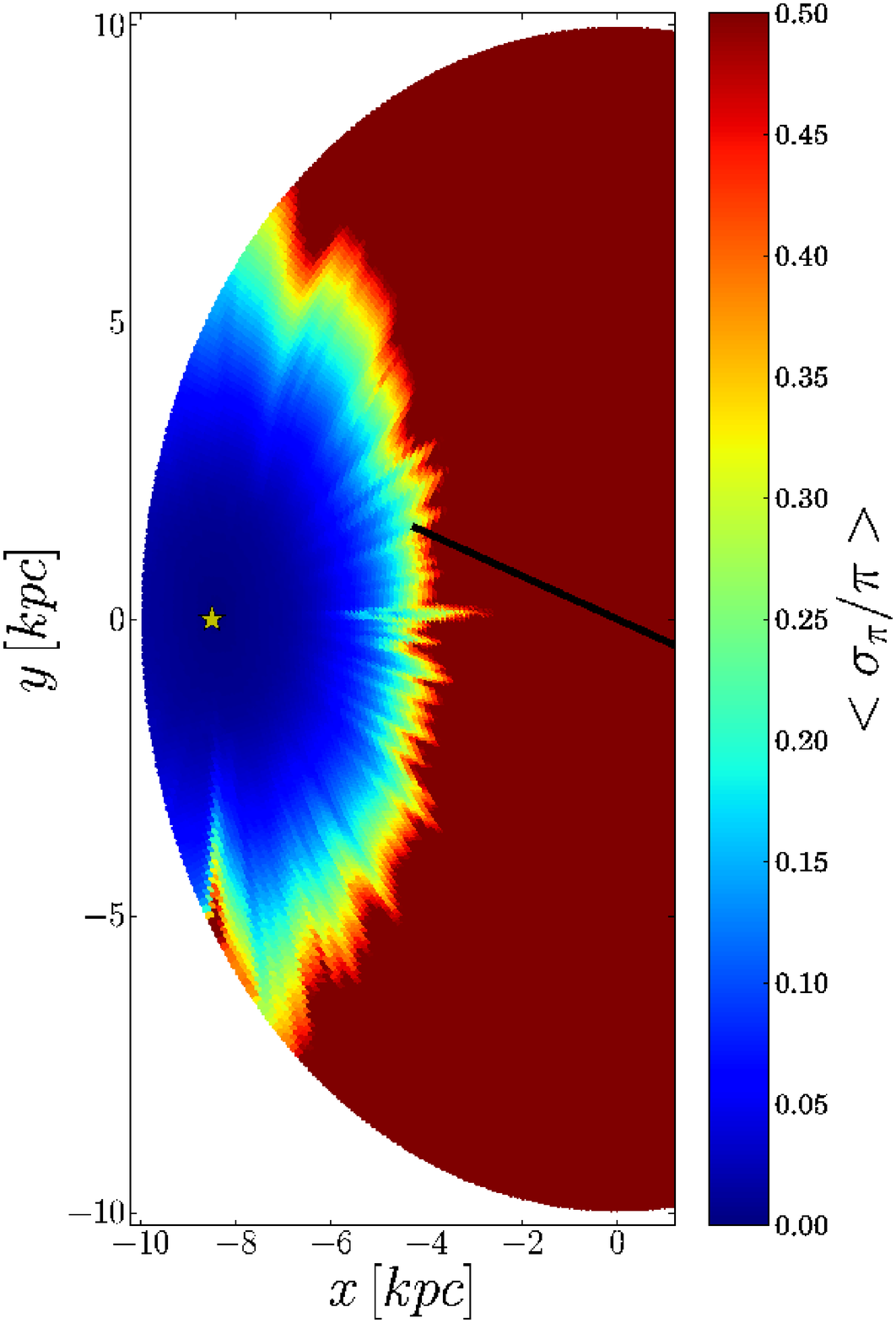}
\includegraphics[scale=0.17]{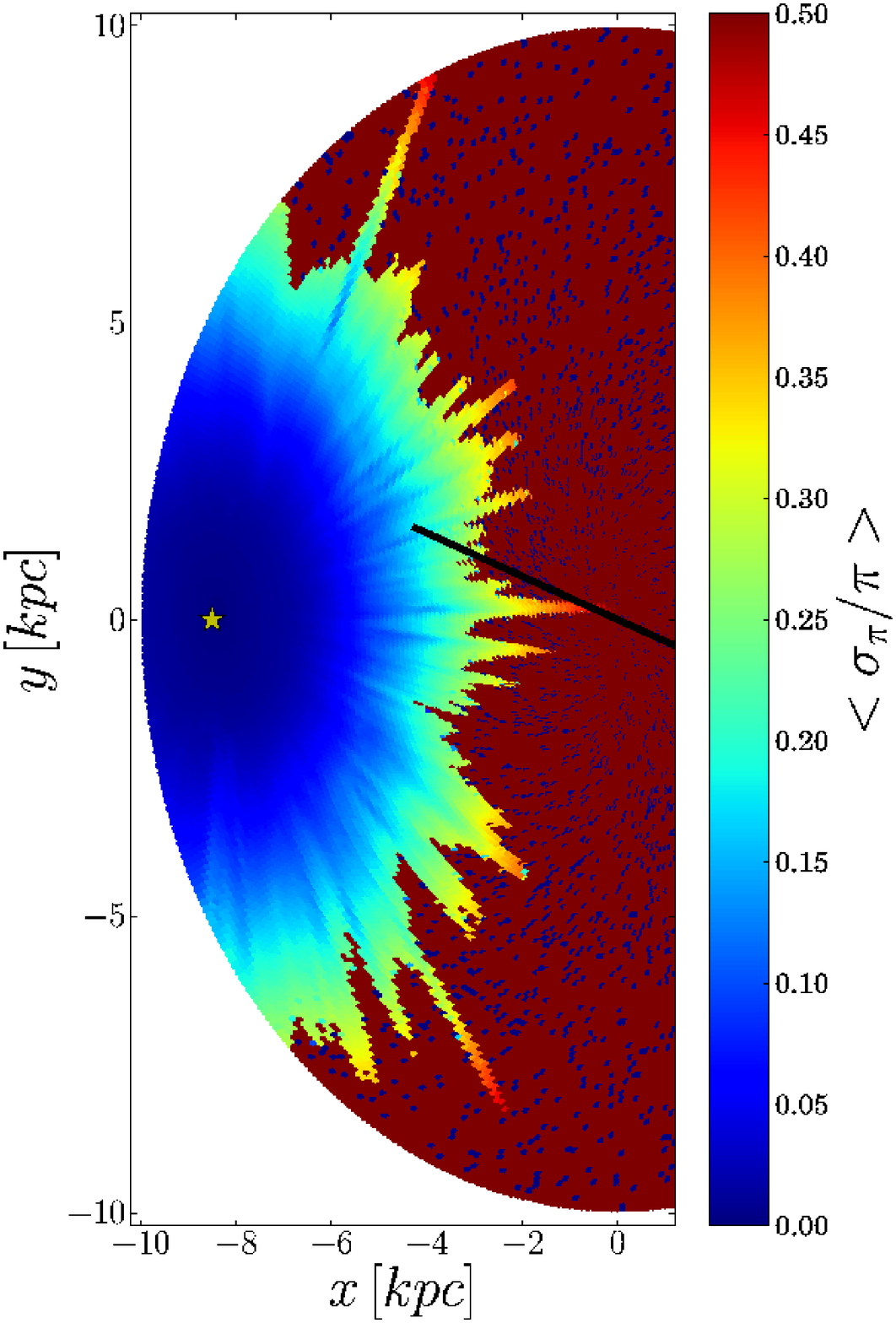}\\
\includegraphics[scale=0.17]{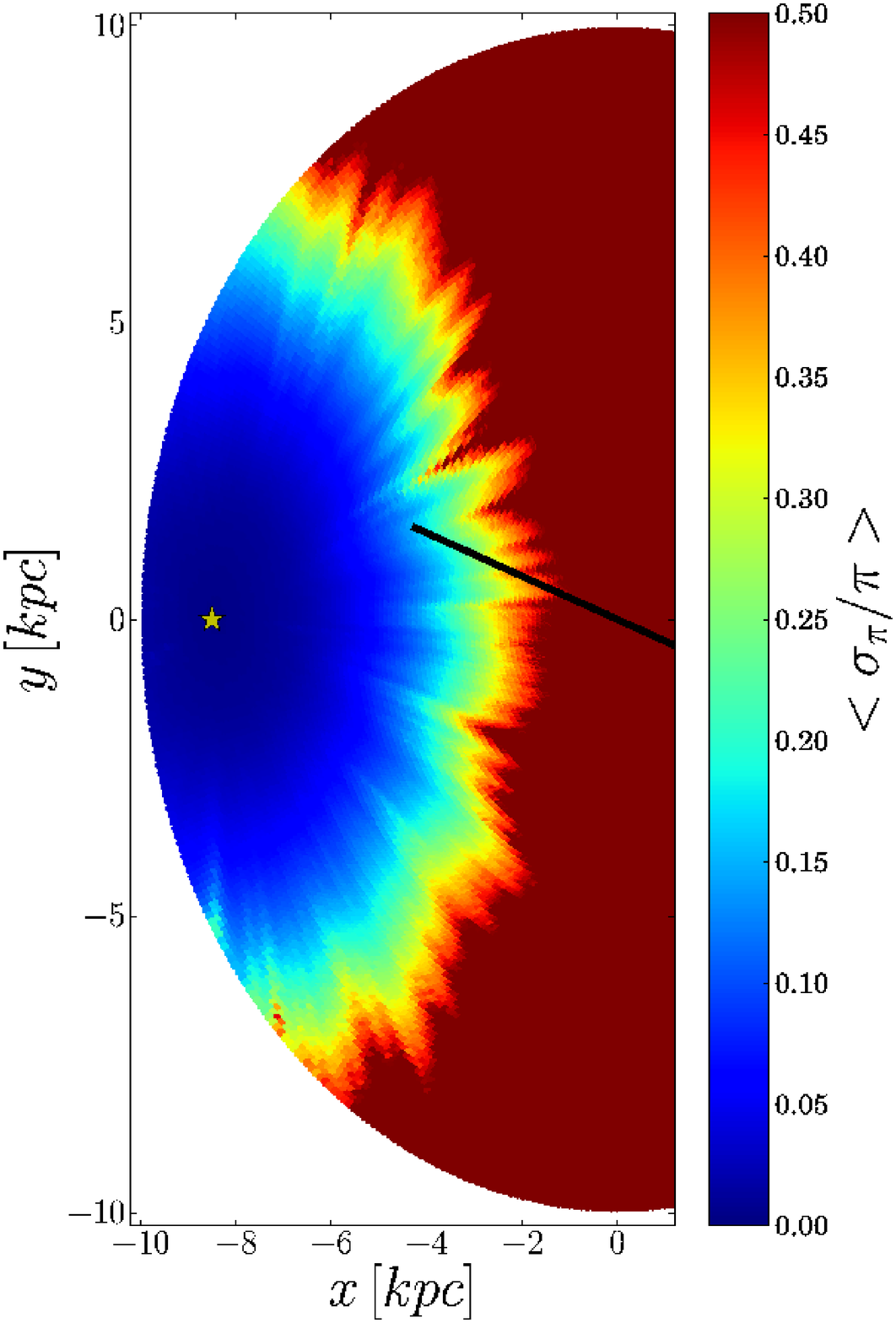}
\includegraphics[scale=0.17]{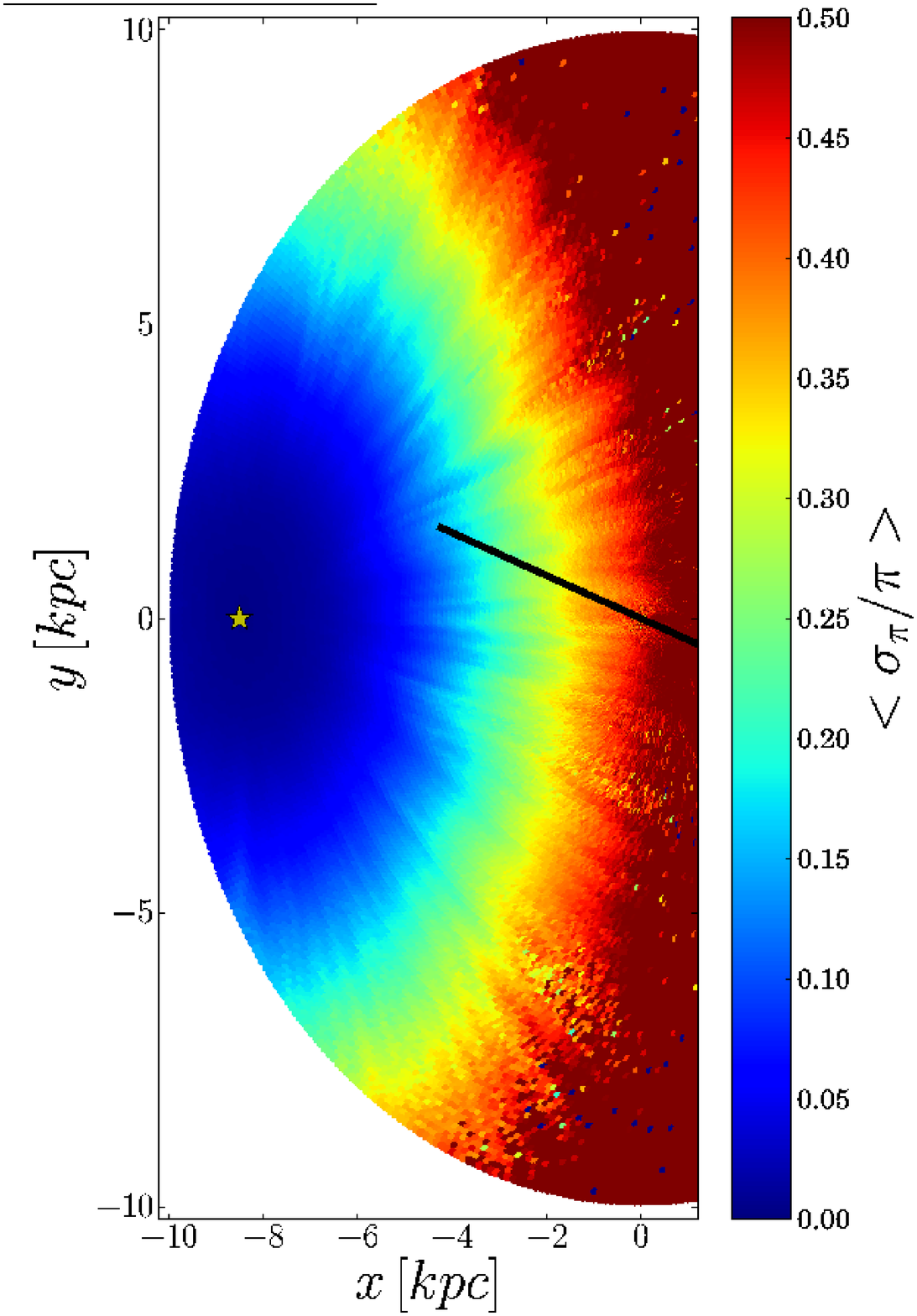}
\end{center}
\caption{
The mean relative error in parallax for the RC-G20 (left) and 
RC-RVS (right) samples for stars closer to the Galactic plane,
$|z|<300\,pc$ (top) and beyond $|z|>300\,pc$ (bottom). Note that the stars are 
distributed using their real distances, i.e. not affected by errors.
The yellow star in all panels shows the position of the Sun, while
the black solid line marks the position and length of the bar.}
\label{fig:error_samples}
\end{figure}

 In this section we study the mean relative 
error in parallax in the RC-G20 and RC-RVS samples. In 
the following plots, we show only the near side of the Galaxy,
that is $x<1$kpc. Due to extinction effects, the vertical distribution
of the particles inside a cylinder perpendicular to the plane is not 
maintained for a given line-of-sight, therefore, 
in Fig.~\ref{fig:error_samples} we 
divide each sample in two, namely particles closer
to the Galactic plane with $|z|<300$pc (top), and particles less 
affected by the extinction, this is, particles such that $|z|>300$pc 
(bottom). As expected, the distribution of the mean relative errors in 
parallax changes through the Galactic plane when we compare the RC-G20 (left) 
and RC-RVS samples (right). For a fixed position in the disc, requiring a 
sample with good radial velocities translates into a 
sample with less number of particles more distributed above 
the plane as seen in the bottom right panel of Fig.~\ref{fig:Nvol_samples}.
When we compare the samples with particles closer to the galactic plane 
(top panels) we see that the RC-RVS sample reaches a given mean relative error
in parallax at larger distance than the RC-G20 sample. This is because, 
at this distance, the stars that remain in the RC-RVS sample are in mean 
brighter than the corresponding RC-G20 sample. As a consequence, the mean
relative error in parallax at this point is smaller. At the same 
heliocentric distance, the RC-G20 sample is fainter in mean and, therefore,
the mean relative error in parallax is larger.
The same argument holds when we compare the sample for stars located at
$|z|>300\,pc$, so less affected by extinction (bottom panels). At the same
heliocentric distance the number of stars per bin of the RC-RVS sample is
significantly less and only the bright stars contribute to the mean, thus, 
it has smaller mean relative error in parallax.

\begin{figure}
\begin{center}
\includegraphics[scale=0.3]{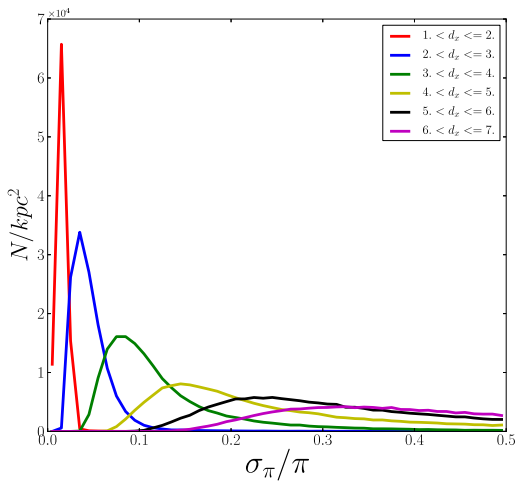}
\end{center}
\caption{
Histograms of the relative error in parallax as a function of the 
distance along the Sun - Galactic Centre line for the RC-G20 sample.}
\label{fig:error_SGC}
\end{figure}

The relative error in parallax for the particular case of the Sun-Galactic 
Centre line is shown in Fig.~\ref{fig:error_SGC} for the RC-G20 sample. 
Each curve represents the number of stars per $kpc^2$ as a function of the 
relative error in parallax for the particles located in the Sun - Galactic 
Centre line. Each colour corresponds to particles that are located at a 
certain distance from the Sun projected on the x-axis, $d_x$, before 
introducing Gaia errors. As expected, particles located close to the Sun's 
position have smaller errors, while the number of particles with higher 
errors increases as we move away from it. Also note that $95\%$ of the 
 RC stars in this line-of-sight will have a relative error in parallax 
less than $20\%$ and we will  have particles with this relative parallax
error up to $5\,kpc$ from the Sun  (yellow line). The relative error 
in parallax for four specific low extinction lines below
the Galactic plane, including the Baade's window, are shown in 
Fig.~\ref{fig:bravafields} in Sect.~\ref{sec:brava}.

\begin{figure}
\begin{center}
\includegraphics[scale=0.18]{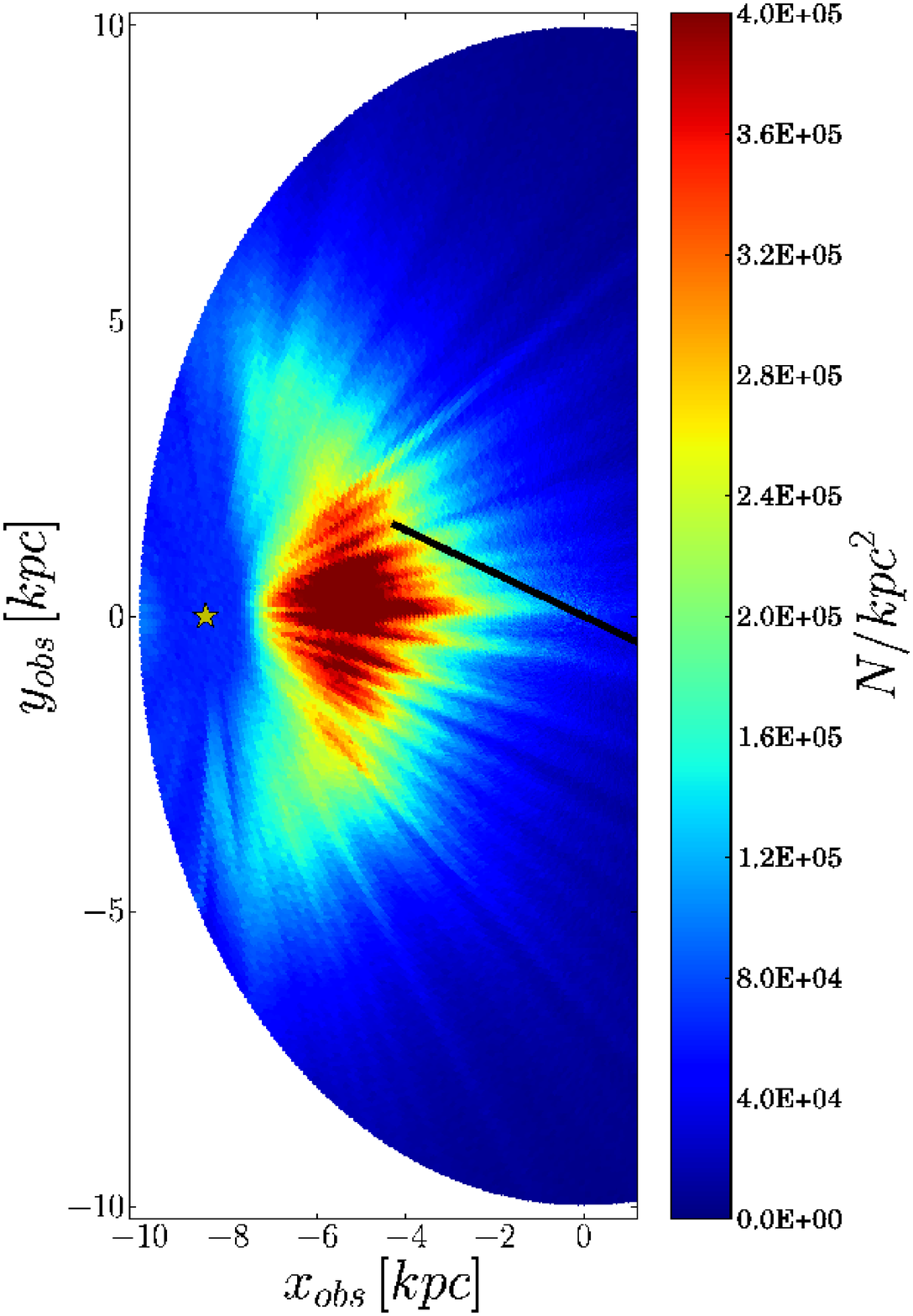}
\includegraphics[scale=0.18]{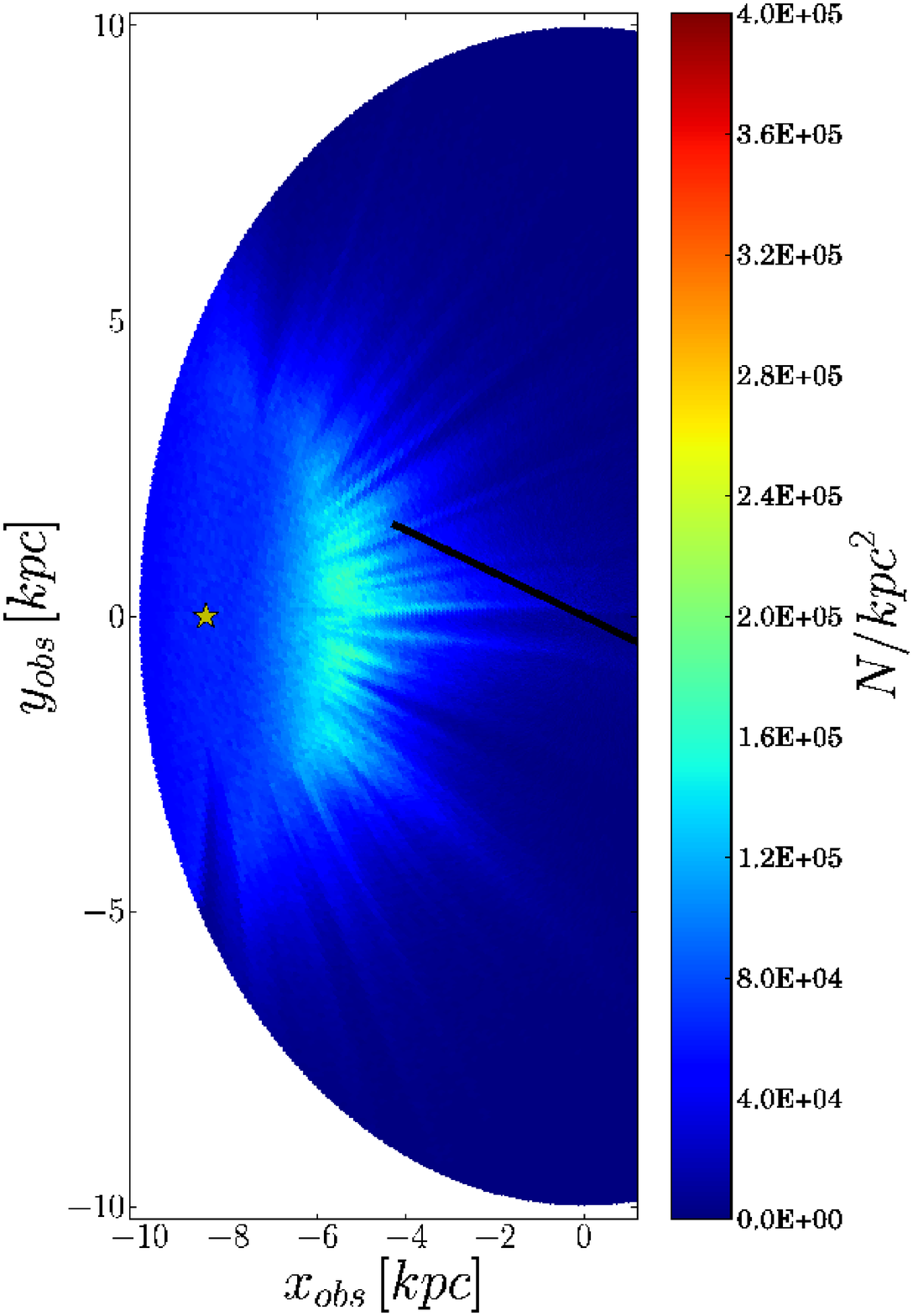}\\
\end{center}
\caption{As in Fig.~\ref{fig:Nvol_samples} for the samples RC-G20-O (left 
panel) and RC-RVS-O (right panel). The stars are distributed using their 
observed distances, i.e. affected by errors. The colour bar is 
different for each sample. The yellow star in both panels 
shows the position of the Sun. The solid black line shows the position and
length of the bar.}
\label{fig:Nvol_obssamples}
\end{figure}

The distribution of errors in parallax shown in Fig.~\ref{fig:error_samples} 
translates in a redistribution of the particles when we add the Gaia errors. 
In Figure~\ref{fig:Nvol_obssamples}, we plot the surface density of the 
RC-G20-O (left) and RC-RVS-O (right) using the observed distances. 
The overdensity due to the bar is blurred and  an accumulation of the
particles takes places towards the Sun so that the central longitudes
look more dense. Furthermore, we have to take 
into account that distances are biased when derived from the observed 
parallaxes. Since 
their relation is non-linear, the value $1/\pi_{obs}$ is a biased estimate
of the true distance \citep{bro97}. The authors propose several methods 
to use astrometric data with minimal biases, such as using stars with
the best relative errors, $\sigma_{\pi_{obs}}/\pi_{obs}<10\%$, though some biases
due to the truncation are still expected, or other methods using all
available information, though they are model dependent \citep[and references 
therein]{bro97}. As discussed in Sect.~\ref{sec:pxpy}, one of the 
suggestions in this work is working directly in 
the space of observables of the catalogue, here, the parallaxes.

 The sky-averaged positions and proper motion relative errors distributions
will follow the same pattern as the mean relative error in parallax presented
here. As discussed in the Gaia Science Performance webpage, direct relations can be used, derived from scanning-law simulations. 

\subsection{Tangential and radial velocities precision distribution}
\label{sec:RCkine}
In this section, we perform a similar analysis of the errors as in 
Sect.~\ref{sec:par_error}. We compute the mean error in tangential velocity 
for both samples, RC-G20 and RC-RVS  and for two subsamples, namely 
one with stars above and below $|z|=300$pc, as in Sect.~\ref{sec:par_error}. 
For RC-RVS, we also compute the mean error in radial velocity. It is essential 
to know how the errors are distributed if we are interested in specific 
science projects, such as analysing the moments of the velocity distribution 
function (Romero-G\'omez et al, in preparation).

\begin{figure}
\begin{center}
\includegraphics[scale=0.17]{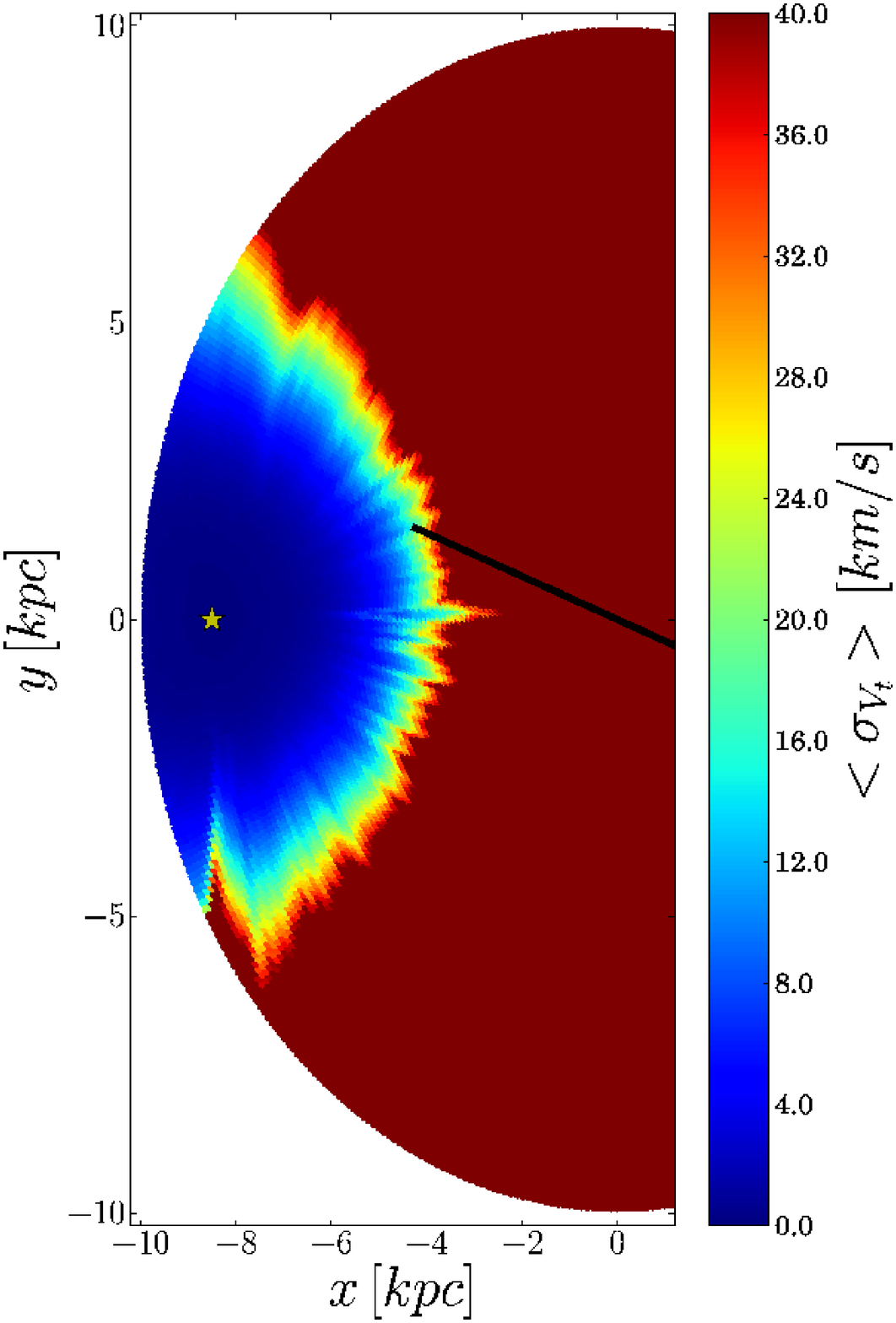}
\includegraphics[scale=0.17]{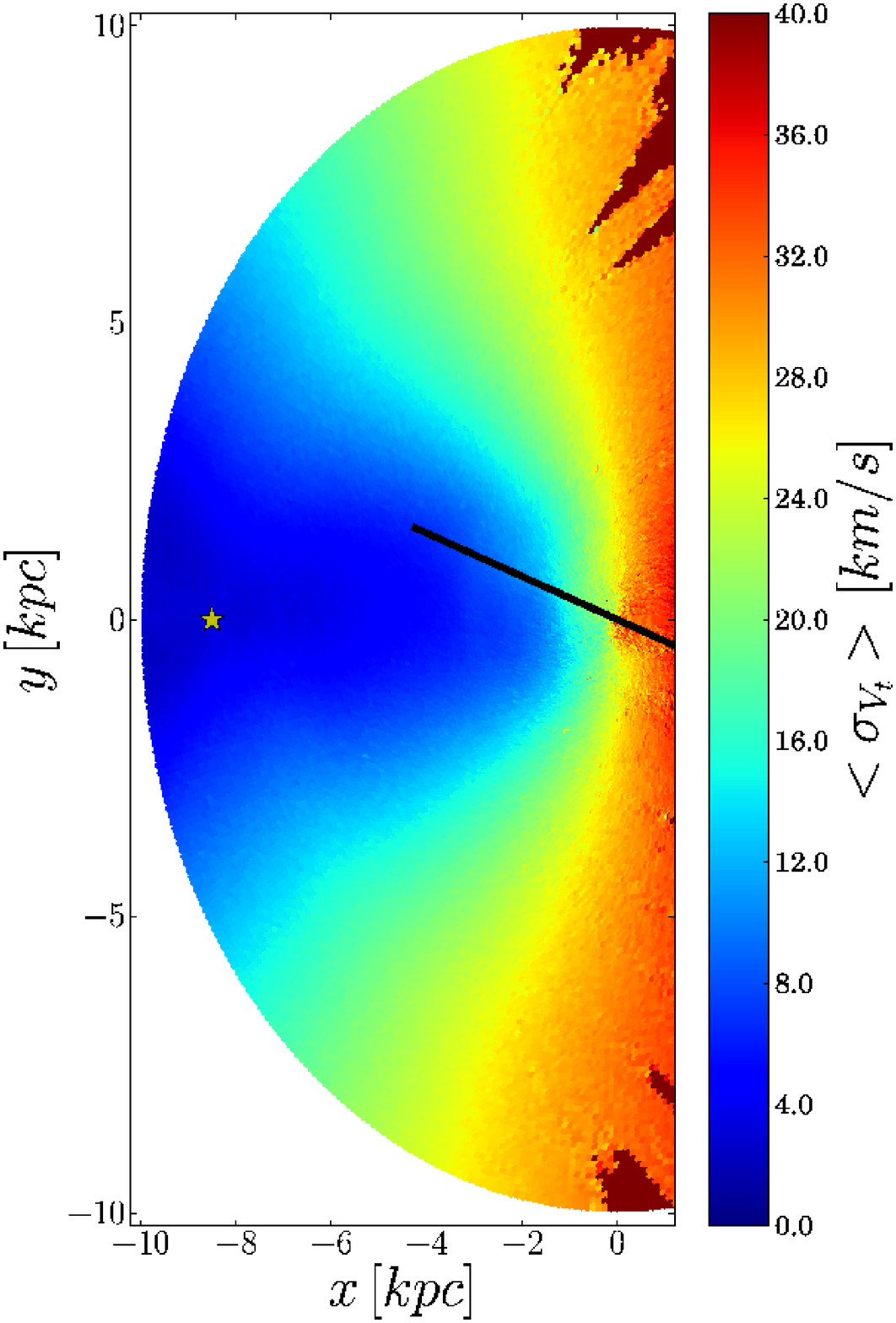}\\
\includegraphics[scale=0.17]{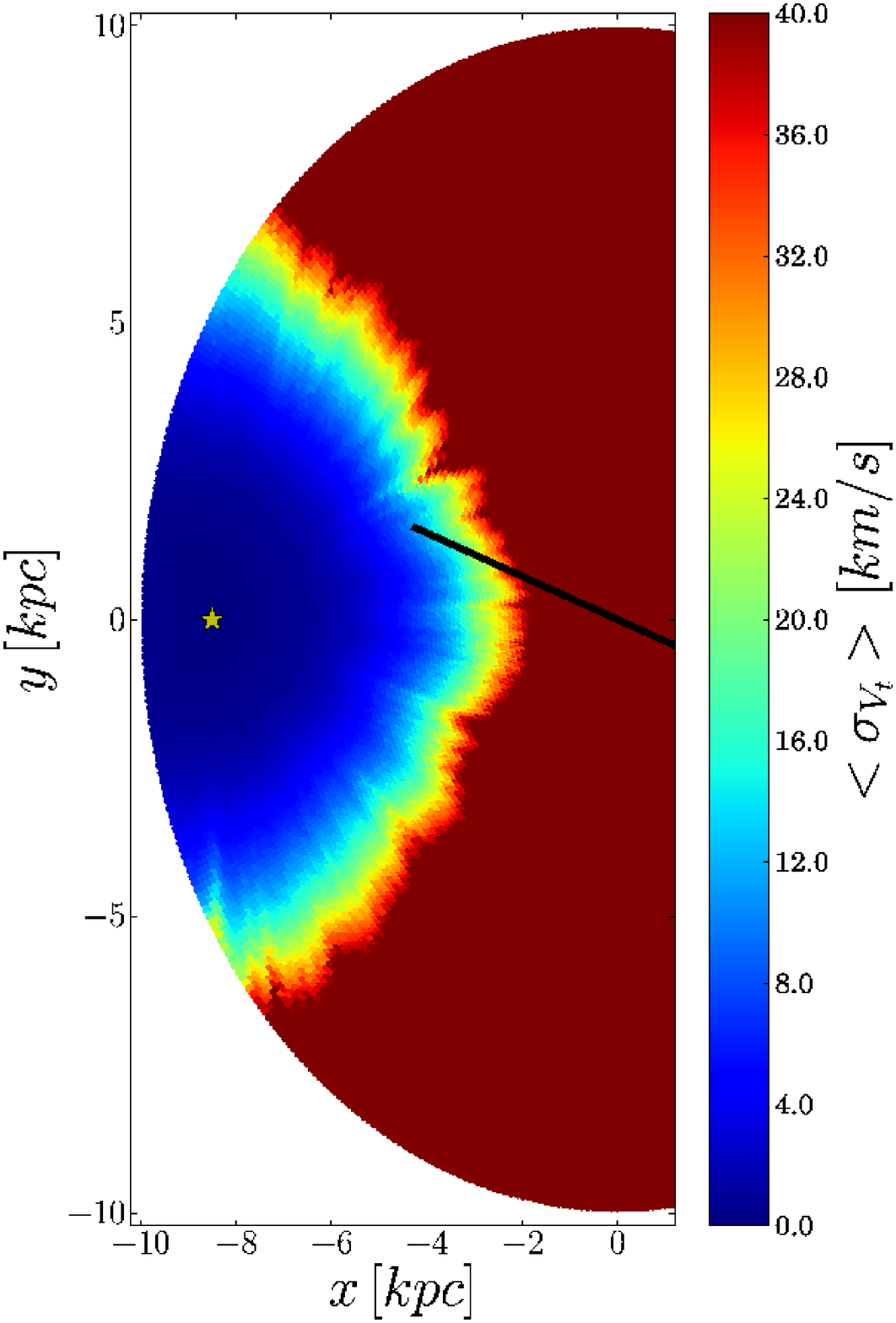}
\includegraphics[scale=0.17]{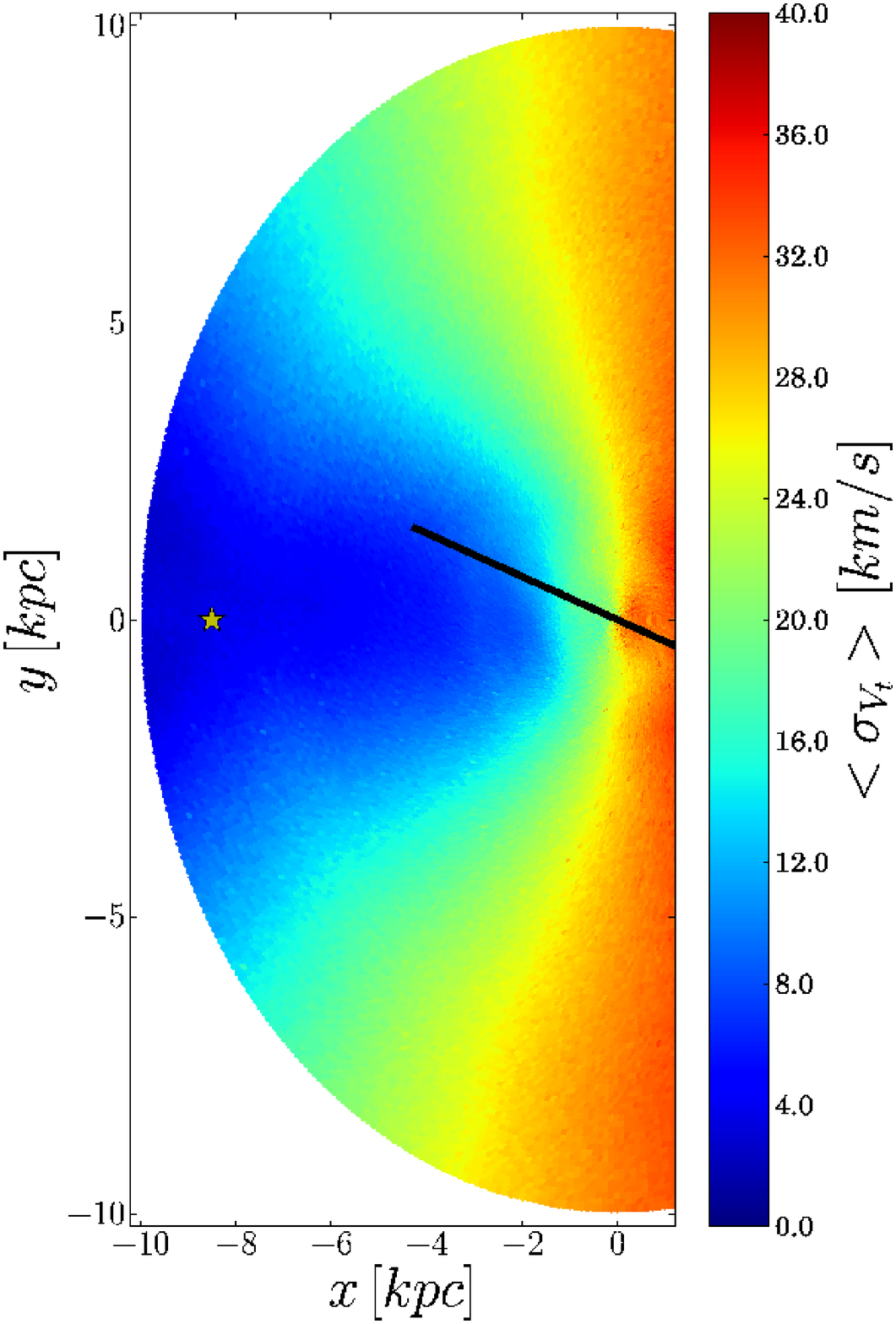}
\end{center}
\caption{Mean errors in tangential velocity. Left panels, for the RC-G20 
sample, that is, using the Gaia error model, and 
the stars distributed using the real distances. Right panels, mean errors in 
tangential velocity for the RC-G20-IR sample, that is, estimating 
the relative error in the distance to be constant and of $10\%$, which 
would correspond if the distances were obtained from the IR photometry. 
 Top 
panels, for stars closer to the Galactic plane $|z|<300$pc. Bottom panels,
for stars above $|z|>300$pc.
The yellow star in both panels shows the position of the Sun. }
\label{fig:kineRC-G20}
\end{figure}

We compute the mean error in tangential velocity on the 
plane in the sky, i.e. $V_t=k\mu d$,  where $\mu$ is the proper motion 
defined as $\mu^2=\mu_{\alpha^*}^2+\mu_{\delta}^2$, and $\mu_{\alpha^*}=\mu_{\alpha}\cdot\cos(\delta)$ and $\mu_{\delta}$ are the proper motions in the right 
ascension and declination, respectively. $d$ is the heliocentric distance 
to the star and 
$k=4.74 \frac{km/s}{``\,yr^{-1}pc}$ for the unit conversion. To compute the
error in tangential velocity, we use, according to the Gaia Science 
Performance webpage, that the mean end-of-mission error in proper motion is
$\sigma_{\mu}=0.526\sigma_{\pi}$, that the error in heliocentric distance
is $\sigma_d=\sigma_{\pi}/\pi^2$. Both errors are taken into account to
derive the error in the tangential velocity, $\sigma_{V_t}$. 
In the left panels of Fig.~\ref{fig:kineRC-G20}, we plot the mean error in 
tangential velocity for the particles in the RC-G20 assuming real distances
 and for the two subsamples above and below $|z|=300$pc. 

This plot suggests that if we want a sample with maximum mean tangential 
velocity error of, for example, $10$\kms, the sample will reach up to 
$3.5$kpc from the Sun  with stars closer to the Galactic plane and
they will reach up to $4.5$kpc for stars above $|z|>300$pc, 
approximately in any direction. If we use IR photometric distances, the coverage
increases up to half of the Galactic bar region in all heights (see right 
panels of Fig.~\ref{fig:kineRC-G20}). As expected, the errors in the end of 
the bar region are much lower when we use IR photometric distances than 
astrometric parallaxes. This justifies the need of IR surveys in the case 
we want to reach deeper in the disc. 

\begin{figure}
\begin{center}
\includegraphics[scale=0.3]{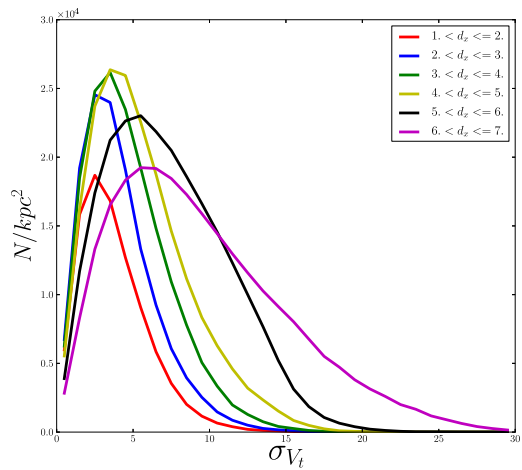}
\end{center}
\caption{Histograms of the error in tangential velocity as a function of 
the distance along the Sun - Galactic Centre line for the RC-G20 sample. }
\label{fig:kineRC-G20-SGC}
\end{figure}

In Fig.~\ref{fig:kineRC-G20-SGC}, we compute the histograms of the error in 
tangential velocity as a function of the distance projected on the x-axis 
to the Galactic Centre in the Sun - Galactic Centre line for the RC-G20 sample. 
 Note how we will be able to have most of the stars with errors less than 
$10$\kms up to $5$kpc from the Sun.

\begin{figure*}
\begin{center}
\includegraphics[scale=0.23]{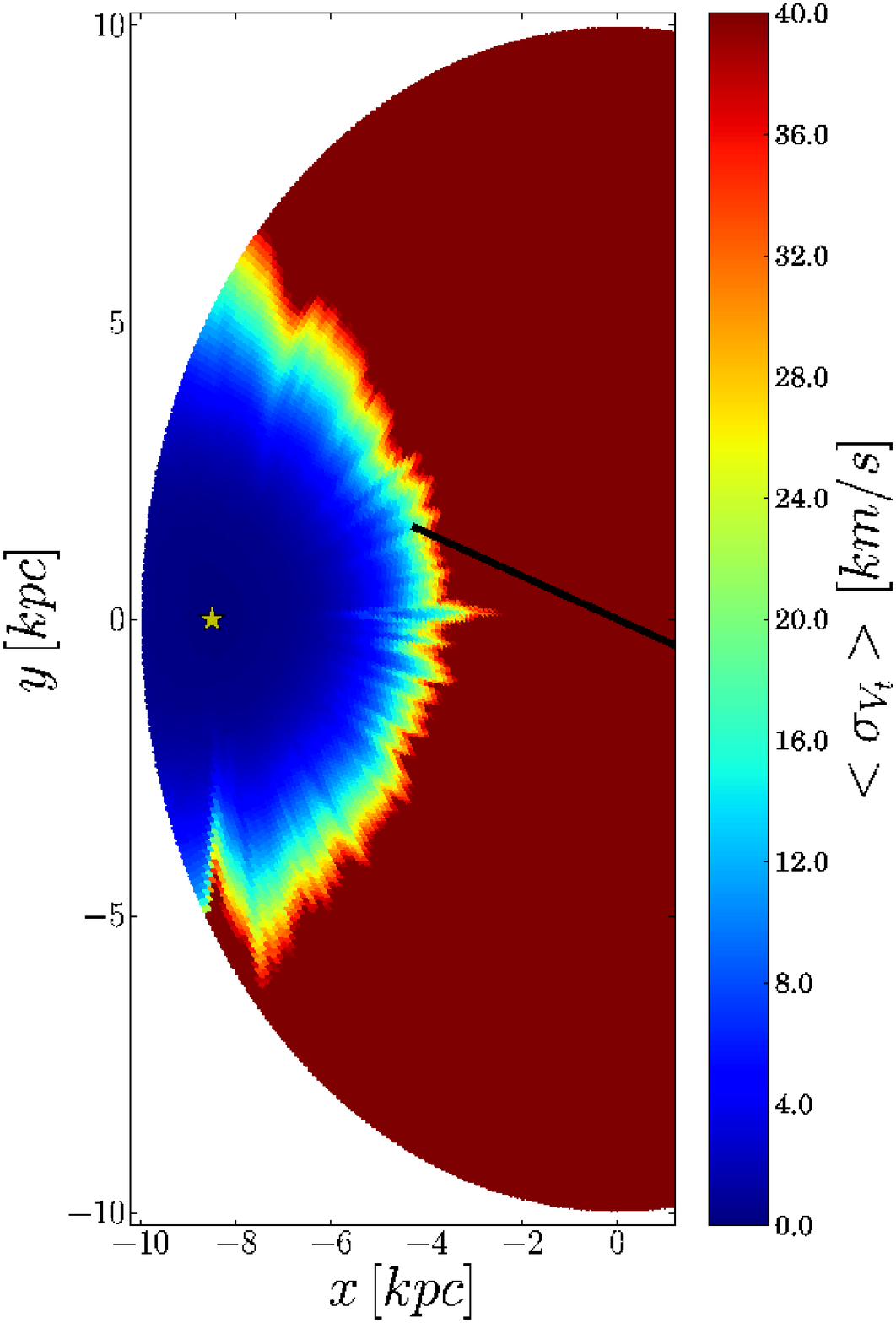}\hspace{0.1cm}
\includegraphics[scale=0.23]{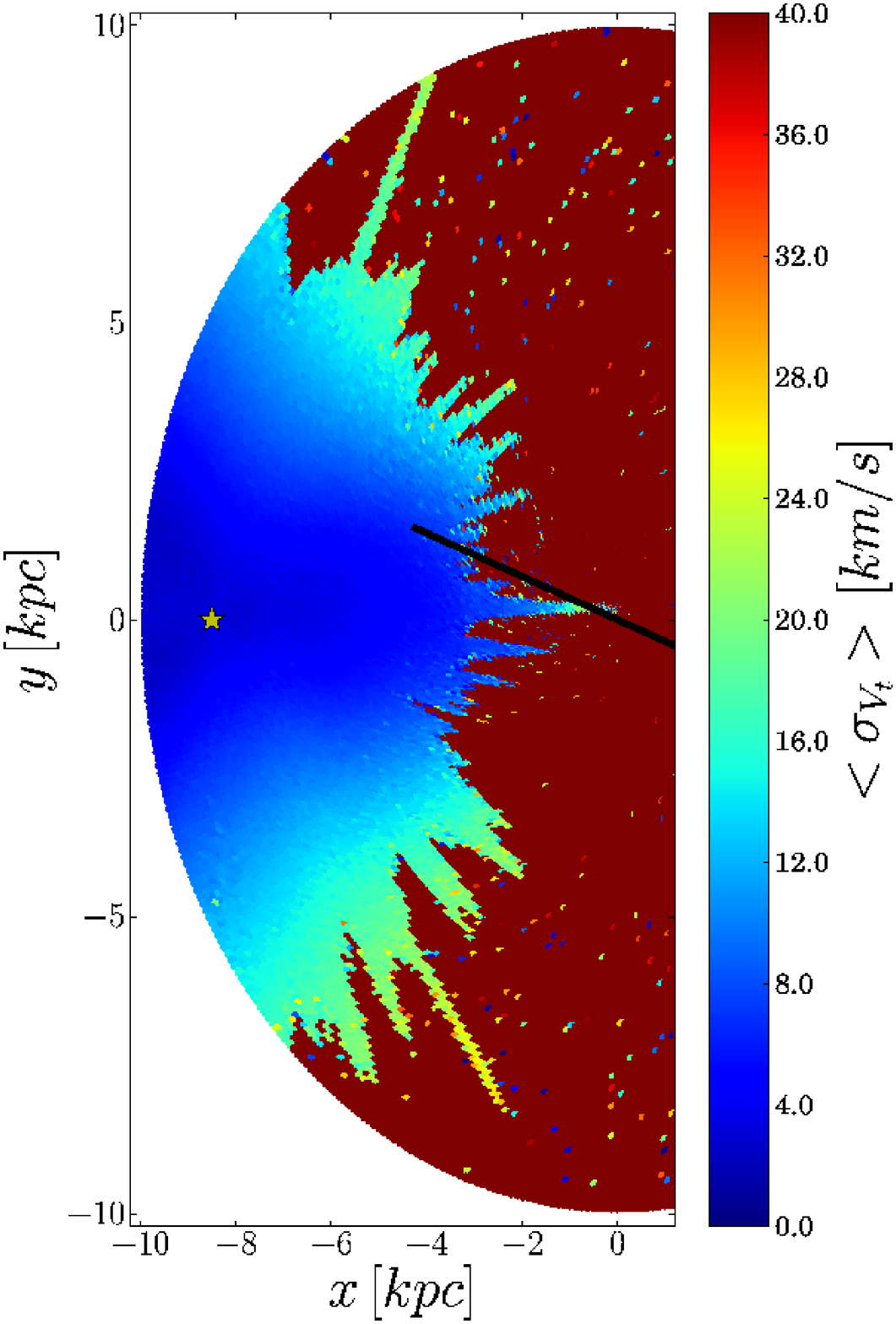}\hspace{0.1cm}
\includegraphics[scale=0.23]{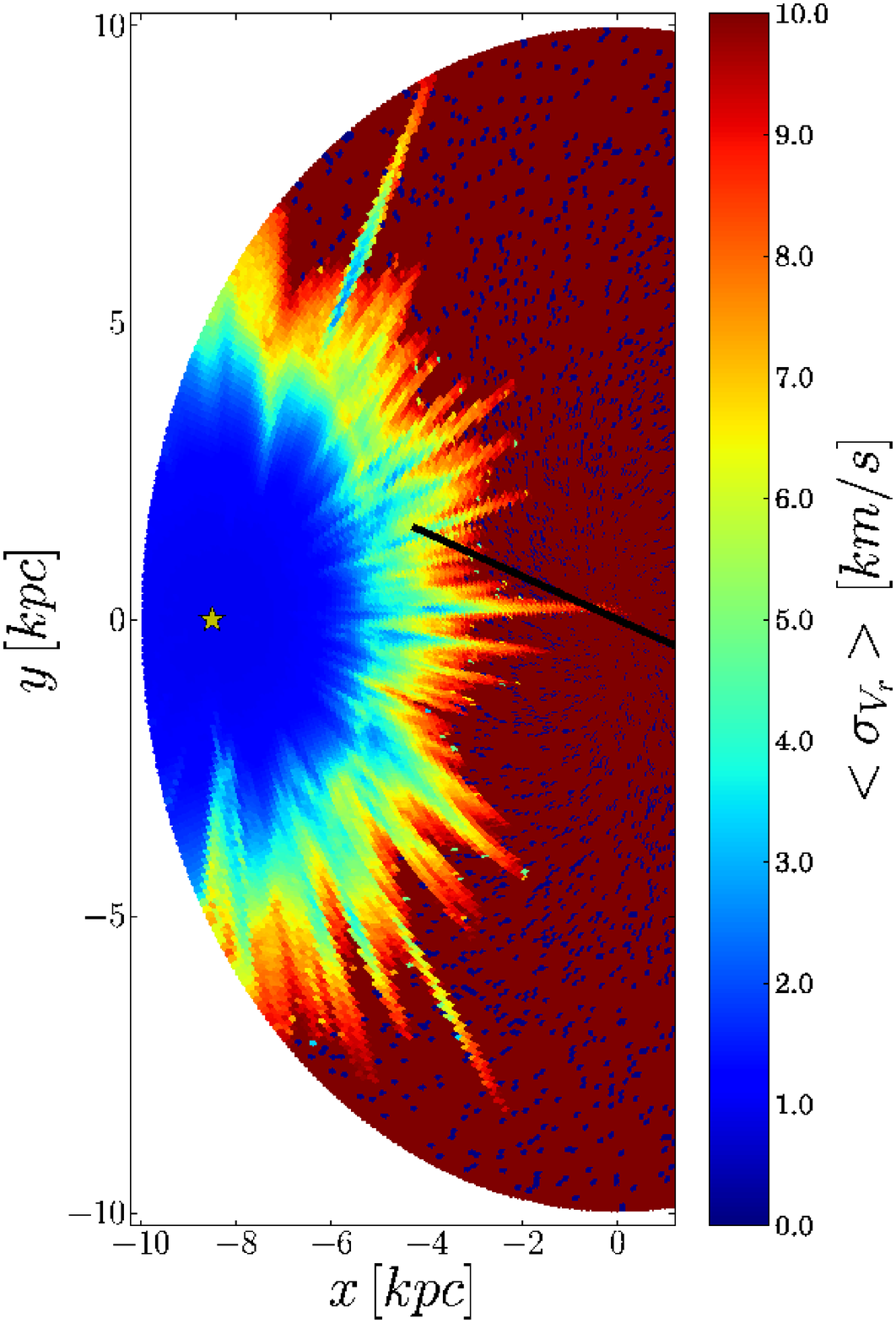}\\
\includegraphics[scale=0.23]{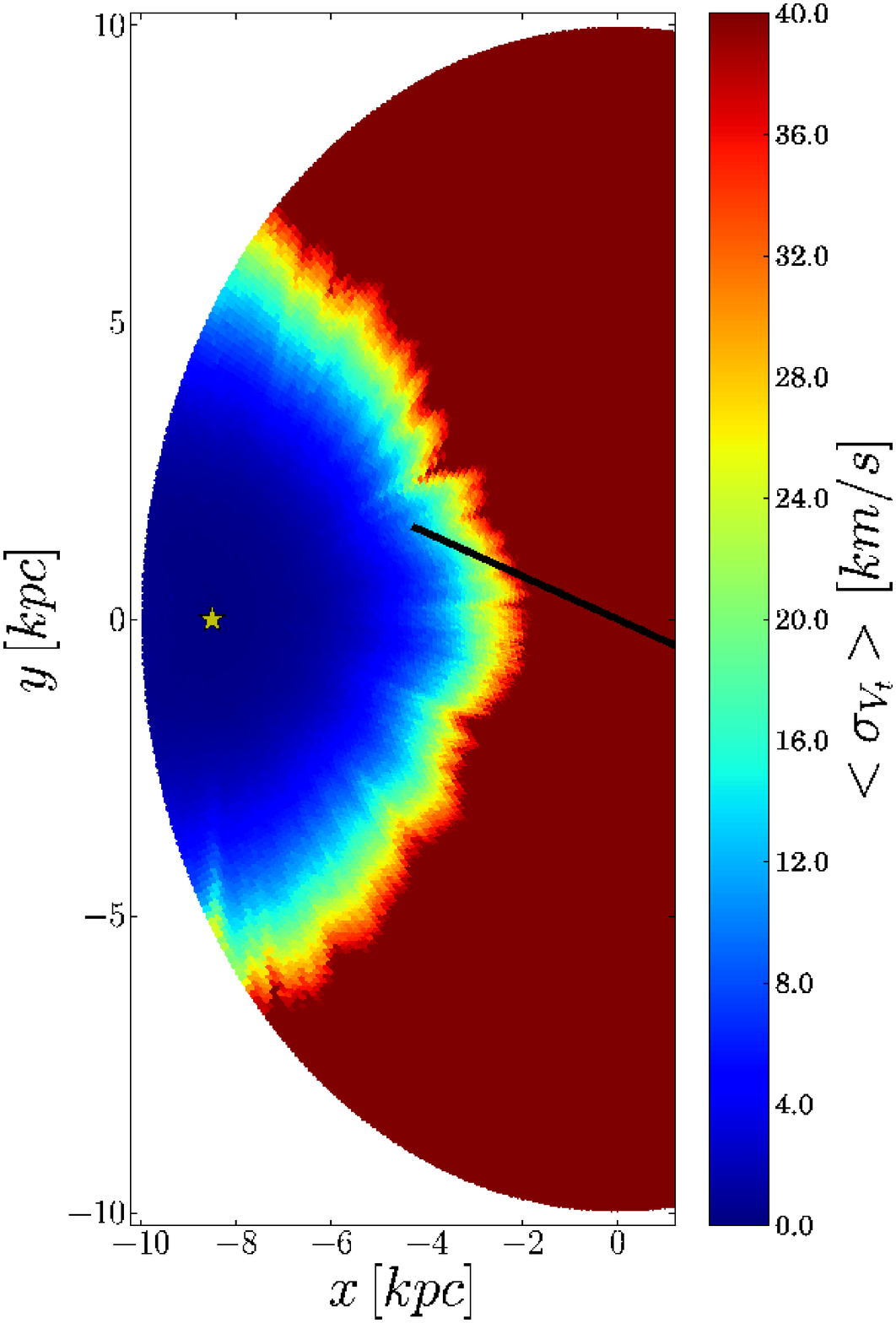}\hspace{0.1cm}
\includegraphics[scale=0.23]{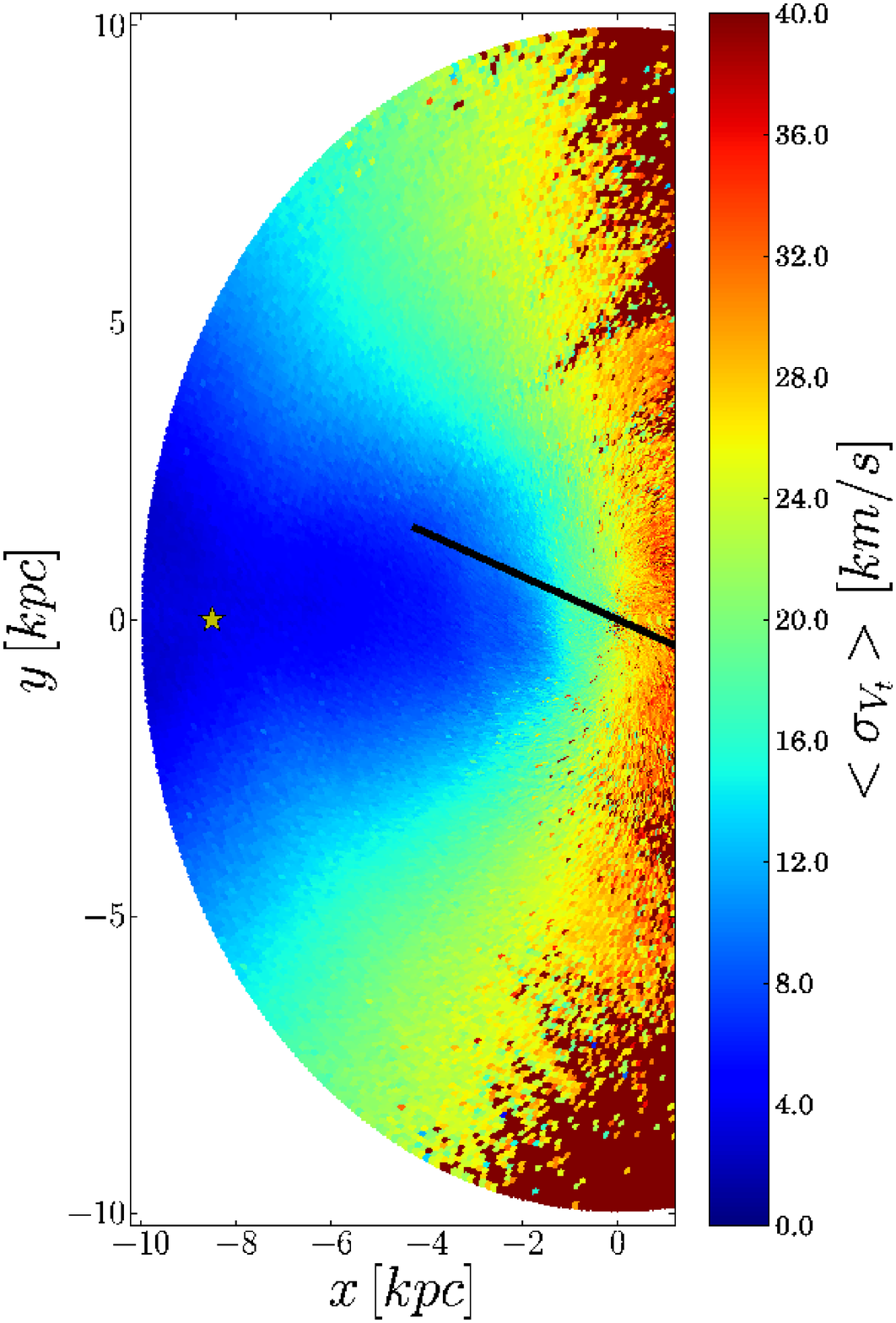}\hspace{0.1cm}
\includegraphics[scale=0.23]{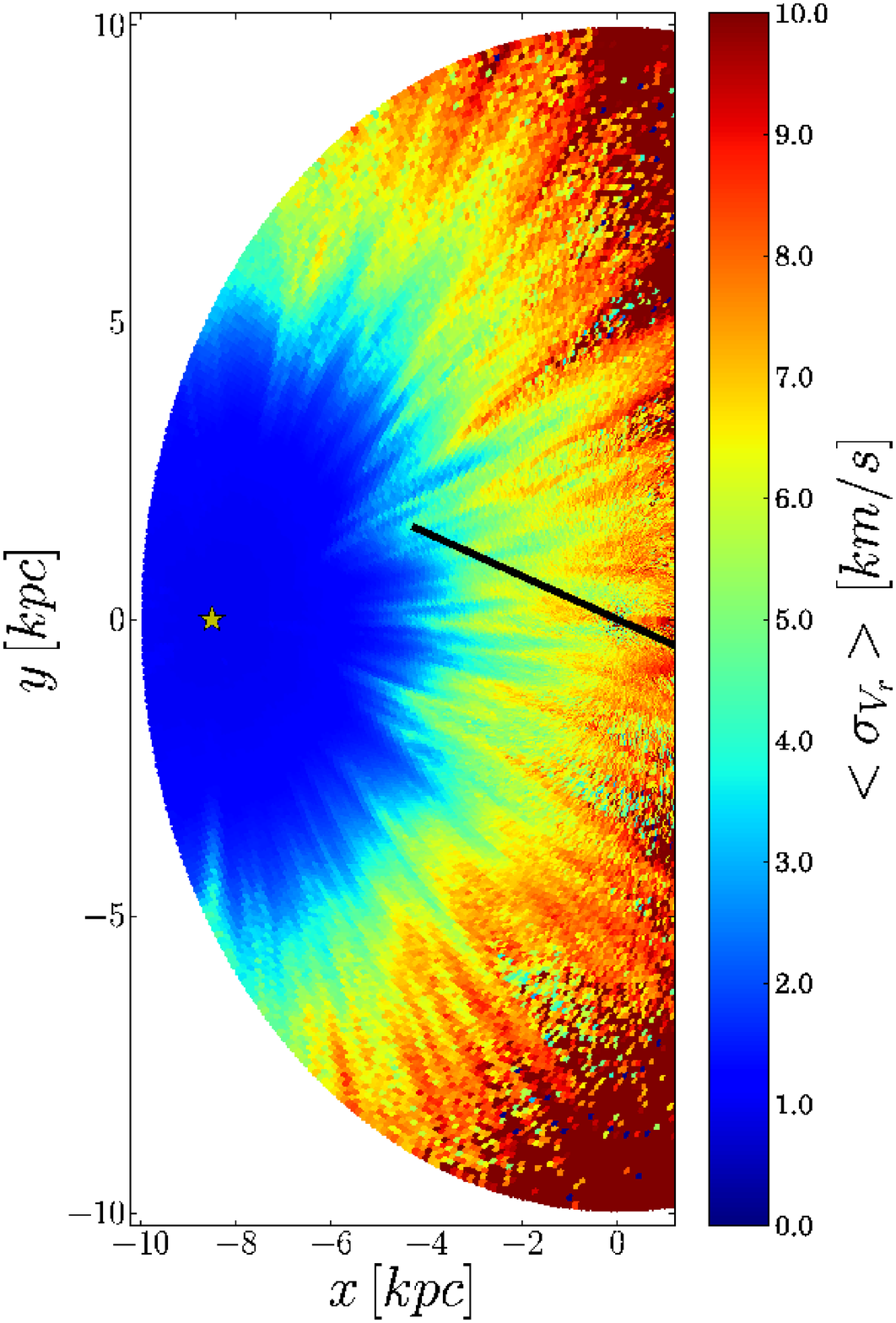}
\end{center}
\caption{Mean errors in tangential and radial velocity in the RC-RVS sample. 
The stars are distributed using the real distances in the regions.
Left and middle: Mean errors in tangential velocity using the Gaia error model
and the IR photometric distance, respectively. Right: Mean errors in radial 
velocity. Top panels: for the subsample containing stars below $|z|<300$pc.
Bottom panels: for the subsample containing stars above $|z|>300$pc.
The yellow star in all panels shows the position of the Sun. }
\label{fig:kineRC-RVS}
\end{figure*}

In Fig.~\ref{fig:kineRC-RVS}, we plot the mean errors in tangential velocity, 
again using both Gaia and photometric distance errors, and the mean errors 
in radial velocity for the RC-RVS sample  and we divide again the sample 
in the particles below $|z|<300$pc (top panels) and above $|z|>300$pc 
(bottom panels). There is a big region,  which is larger if 
the stars are outside the Galactic plane, for which 
the mean errors in tangential velocity obtained using Gaia, which includes 
the end of the bar region, are less than $15$\kms. But this region 
increases including almost half of the bar region, when using photometric 
distances. The mean errors in radial velocity in the bar region and outside
the Galactic plane are in the range $4-6$\kms, while within the Gaia sphere 
the error is less than $4$\kms, for all heights.

In Table~\ref{tab:errors}, we give the mean errors in two specific 
polar regions of $1\,kpc\times 10\gr$, namely the regions near the 
end of the Long bar (hereafter LB region) and the end of the Boxy/bulge bar 
(hereafter BB region), for both the RC-G20 and RC-RVS samples and the two
cuts in height. The galactocentric cylindrical coordinates of these points 
are given in the caption of Table~\ref{tab:errors}. Note that the mean error in 
tangential velocity can be reduced by a factor 5 in some of these regions in 
the disc when using the IR photometric distances.

\begin{table}
\begin{center}
\begin{tabular}{ccccc}
\multicolumn{5}{c}{BB region}\\
\hline
\hline
  & \multicolumn{2}{c}{RC-G20} & \multicolumn{2}{c}{RC-RVS}  \\
\hline
  & $|z|<0.3$ &  $|z|>0.3$ & $|z|<0.3$ & $|z|>0.3$ \\
\hline
$<\sigma_{V_t}>(Gaia)$ & $>50$ & 25.8 & 24.9 & 20.9  \\
\hline
$<\sigma_{V_t}>(IR)$  & 8.6 & 9.2 & 8.4 & 9.1  \\
\hline
$<\sigma_{V_r}>$ & --- & --- & 9.5 & 5.0 \\
\hline
\multicolumn{5}{|c|}{LB region}\\
\hline
\hline
  & \multicolumn{2}{|c|}{RC-G20} & \multicolumn{2}{|c|}{RC-RVS}  \\
\hline
  & $|z|<0.3$ & $|z|>0.3$ & $|z|<0.3$ & $|z|>0.3$ \\
\hline
$<\sigma_{V_t}>(Gaia)$ & 19.6 & 11.5 & 12.8 & 11.4  \\
\hline
$<\sigma_{V_t}>(IR)$  & 6.7 & 7.1 & 6.8 & 7.1  \\
\hline
$<\sigma_{V_r}>$ & --- & --- & 6.3 & 4.4 \\
\hline
\end{tabular}
\caption{Mean errors in tangential and radial velocity for the RC-G20 and
RC-RVS sample in the BB and LB regions and for different cuts in height. 
The region corresponding to LB is $(R,\theta)=(4.5kpc,160\gr)$, and
the one corresponding to BB is $(R,\theta)=(3.1kpc,160\gr)$.
The azimuthal angle is defined positively in the counter-clockwise 
direction from the x-positive axis. Units are kpc and \kms.}
\label{tab:errors}
\end{center}
\end{table}

\subsection{The vertical distribution at the near end of the bar}
In this section we focus on the vertical distribution of stars in the
BB and LB regions, that is, the two regions near the end of the Boxy/bulge 
and Long bar, respectively.

\begin{figure}
\begin{center}
\includegraphics[scale=0.4]{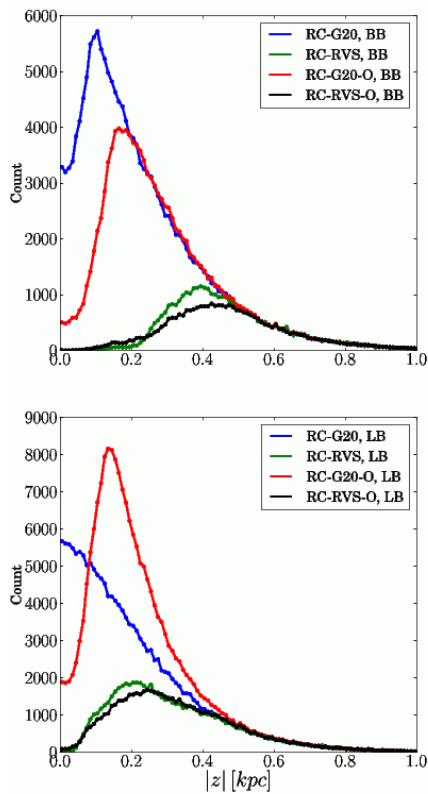}\\
\end{center}
\caption{The vertical distribution of stars in the BB (top panel) and 
LB (bottom panel) regions for the RC-G20 (blue), RC-G20-O (red), RC-RVS 
(green) and RC-RVS-O (black) samples. }
\label{fig:N_z}
\end{figure}

Figure~\ref{fig:N_z} shows how the particles in the BB (top) and LB (bottom) 
regions are distributed in $| z|$ for the RC-G20, RC-G20-O, RC-RVS and 
RC-RVS-O. 
When we consider the RC-G20 sample without Gaia errors (blue lines), the 
number of particles changes depending on the position in the galactic disc. 
In the LB region, the number of particles decreases with height. However, 
in the BB region, the extinction increases and the number of particles 
decreases close to the galactic plane, until about $100$pc, where again
the number of particles decreases with height. The effect of the Gaia
errors depends again on the region. In the BB region, the amount of particles 
close to the galactic plane is much less, until about $100$pc, while
for the LB region, we lose particles in the first $100$pc, but there are
more near $200$pc. This redistribution of particles in height is basically 
an effect of the distance error (see Fig.~\ref{fig:Nvol_obssamples}).

The difference between the RC-RVS sample (green) and the RC-RVS-O sample 
(black) is not so evident, because requiring a high quality 
sample in terms of radial velocities also translates into a high quality 
sample in terms of parallaxes, as seen in Fig.~\ref{fig:error_samples}. 
Therefore, the positions are not so affected by errors. 
There are no particles in the Galactic plane neither in the BB or LB regions. 
However, the number of particles increases up to $200\,pc$ in the 
case of the LB region, while in the BB region, there is still almost no 
particle at this height. 

\subsection{Below the Galactic Plane}
\label{sec:brava}
\begin{figure}
\begin{center}
\includegraphics[scale=0.4]{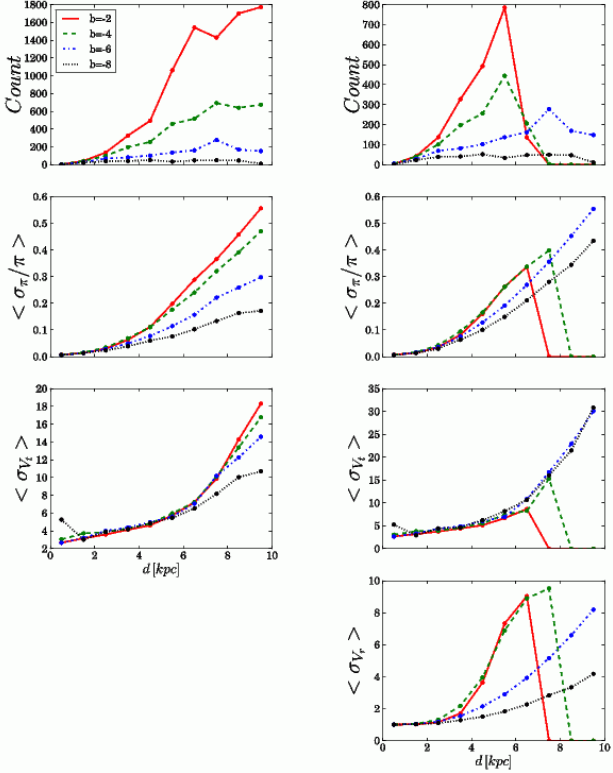}
\end{center}
\caption{Number of particles and errors as a function of  real 
heliocentric 
distance in four BRAVA fields along the $l=0$ line-of-sight, namely $b=-2$ 
(red solid line), $b=-4\gr$ (green dashed line), $b=-6\gr$ (blue dot-dashed 
line) and $b=-8\gr$ (black dotted line) for the RC-G20 sample (left column) and
RC-RVS sample (right column). First row: histogram of the number of 
particles. Second row: mean relative errors in parallax. Third row: mean
errors in tangential velocity (units: \kms). Fourth row: mean errors in 
radial velocity (units: \kms). We do not plot the mean errors in radial 
velocities for the RC-G20 sample because not all particles will have valid 
radial velocities. }
\label{fig:bravafields}
\end{figure}

Other interesting regions in the Galactic disc 
are the ones in the Bulge regions below the Galactic plane. These fields
are interesting for the lower extinction and because spectroscopic
data is being obtained by the BRAVA survey\footnote{BRAVA project website:\\ 
\texttt{http://brava.astro.ucla.edu/index.htm}}. In Fig.~\ref{fig:bravafields},
we plot for both RC-G20 (left panels) and RC-RVS (right panels), information
regarding four of the fields in the $l=0$ line-of-sight, namely 
$b=-2,\,-4,\,-6,\,-8\gr$, as a function of the 
heliocentric distance, binned every $1$kpc. From top to bottom, we
plot the histogram of the number of particles, the mean relative error in
parallax, the mean error in tangential velocity, and
the mean error in radial velocity. The number of RC stars detected by Gaia
can complement the number of M giants in the BRAVA fields. As expected,
the number of particles decrease when decreasing the galactic latitude,
and when $b=-8\gr$, the samples almost run out of particles.
Due to the configuration of the ellipsoid bars imposed (see 
Fig.~\ref{fig:elipse}), at an heliocentric distance of $6\,kpc$ we reach 
the Galactic bar at $l=0$. For the RC-G20 sample, the mean relative error 
in parallax will be of $20-30\%$ at $b=-2\gr$, for stars with an heliocentric 
distance of $6\,kpc$, and even less far below the plane. The mean error in 
tangential velocity is about $5$\kms in all the latitude bins at $6\,kpc$.  

If we analyse the sample with good radial velocities, RC-RVS (right column
of Fig.~\ref{fig:bravafields}), we observe that for the
fields closer to the Galactic plane ($b=-2,-4$), the sample runs out of
particles at about $6$kpc,  due to the imposed extinction law, while
for $b=-6$ the density is approximately constant, but for $b=-8$ it decreases
again. The mean relative error in parallax and 
the mean error in velocity improves with respect to the RC-G20 sample. Using 
the pre-commissioning error model for the radial velocity error, we obtain 
mean errors in radial velocities of maximum $5$\kms, that is, improving 
the errors given in the BRAVA surveys for this field (about $10$\kms) 
\citep{ric07,kun12}.

\section{Characterising the Galactic bar}
\label{sec:internal}
The goal of this section is to provide first insights on the capabilities of
Gaia RC population to recover the value imposed for the azimuthal angle of the
Galactic bar. As mentioned above, the heliocentric distances are biased when 
computed from trigonometric parallaxes. Therefore, in Sect.~\ref{sec:pxpy} 
and for the first time, we introduce the space of Gaia observables 
parallax-galactic longitude. We identify the stellar overdensity distribution
in this plane using the full sample RC-all. The complexity arisen from the 
introduction of the interstellar absorption is well described when using the  
RC-G20 and RC-G20-O samples. As a second attempt, in Sect.~\ref{sec:bar}, we 
introduce additional information coming from the IR (i.e. the RC-G20-IR sample)
to try to recover the angular orientation of the bar. In this case, the
observable is the distance, thus, we show the results in the configuration 
space, i.e. the galactocentric $(x,y)$ plane.

\subsection{The Gaia space of observables}
\label{sec:pxpy}
\begin{figure}
\begin{center}
\includegraphics[scale=0.55]{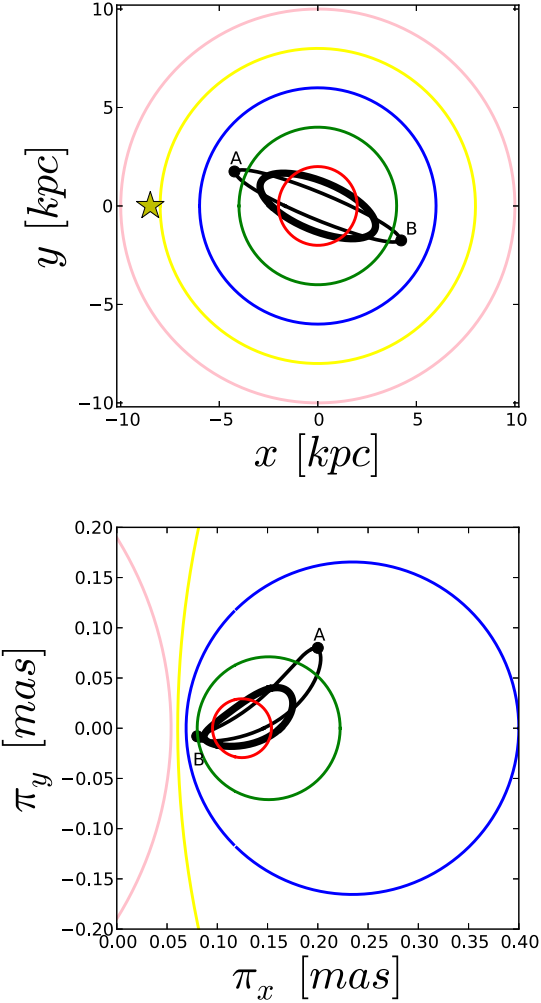}\\
\end{center}
\caption{Mapping structures to the Gaia observable space. Top panel: 
structures in the $(x,y)$ configuration space. We plot in thick (thin) black 
lines the ellipses corresponding to the Boxy/bulge (Long) bar and 
galactocentric circles at 2(red), 4(green), 6(blue), 8(yellow) and 10(pink) kpc.
The star symbol show the position of the Sun. Bottom panel: the same structures 
mapped in the ($(\pi_x,\pi_y)$) plane. }
\label{fig:elipse}
\end{figure}

The space of Gaia observables is defined as $(\pi_x,\pi_y)$, i.e. the converted 
$(\pi,l)$ in cartesian coordinates. An illustration of the coordinate
transformation is presented in Fig.~\ref{fig:elipse}. In the top panel, we 
present two ellipsoidal Ferrers bars: the Long (black thin) and the boxy/bulge 
(black thick) bars, centred at the origin of coordinates, the Galactic Centre, 
and tilted $20\gr$ away from the Sun-Galactic Centre line. They have the same 
parameters as described in Sect.~\ref{sec:model}. Note that the ellipsoids 
correspond to the edges of the Ferrers bars and that the density decreases 
inhomogeneously from the Galactic Centre. The star symbol shows the Sun's 
position located at $8.5$kpc on the negative x-axis. We also add galactocentric 
circles at radii $2$ (red), $4$ (green), $6$ (blue), $8$ (yellow) and $10$ 
(pink) kpc for clarity. The bottom panel shows the same ellipsoidal bars and
circles, but in the observable space. Although they come out in an odd 
looking shape, a close inspection of the labelled points A and B at the 
opposite ends of the Long bar can help understanding the transformation, 
knowing that large distance corresponds to small parallax and vice versa. 
Also note how the galactocentric circles are transformed in the observable 
space into non-concentric circles.

\begin{figure}
\begin{center}
\includegraphics[scale=0.23]{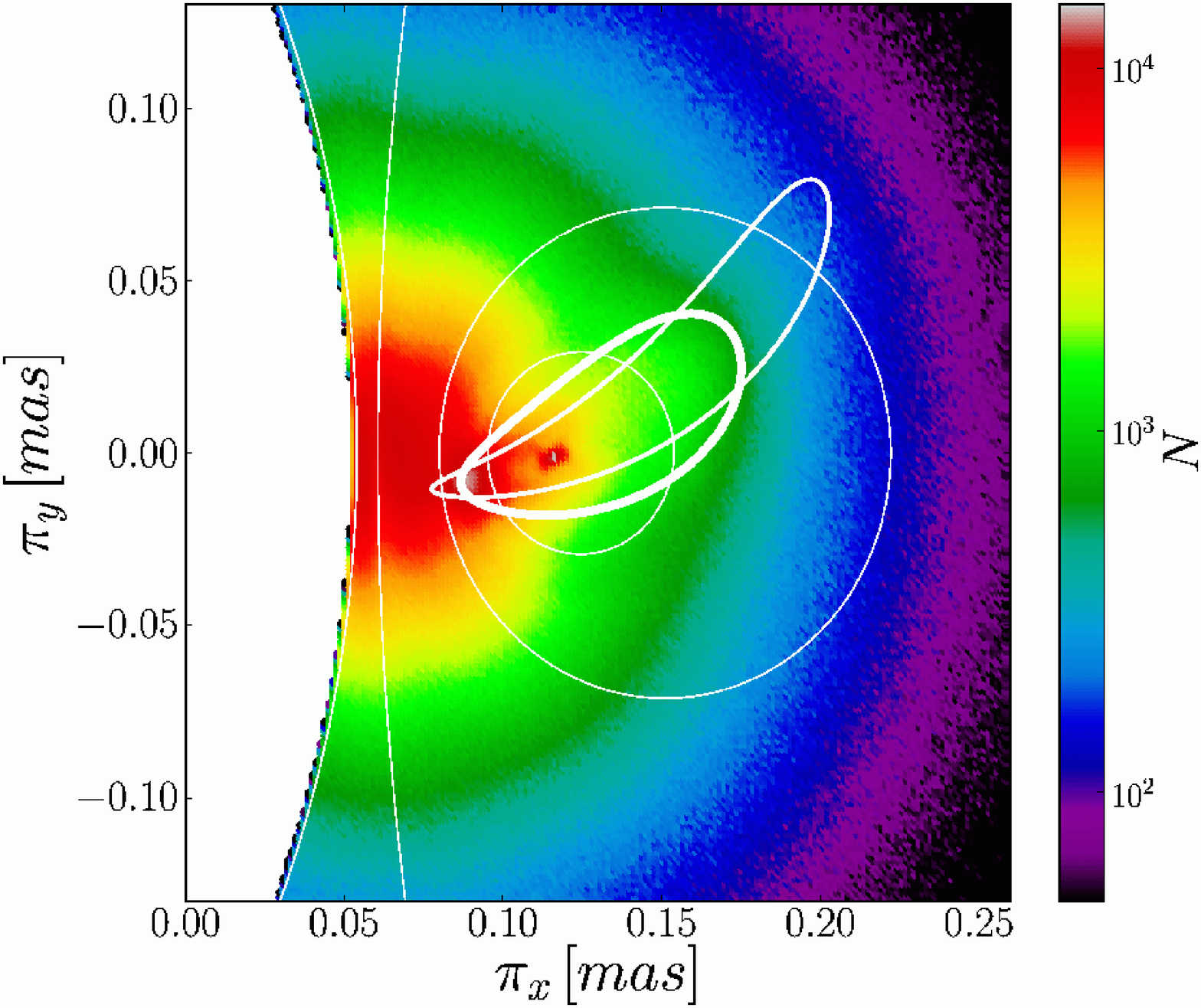}\\
\includegraphics[scale=0.23]{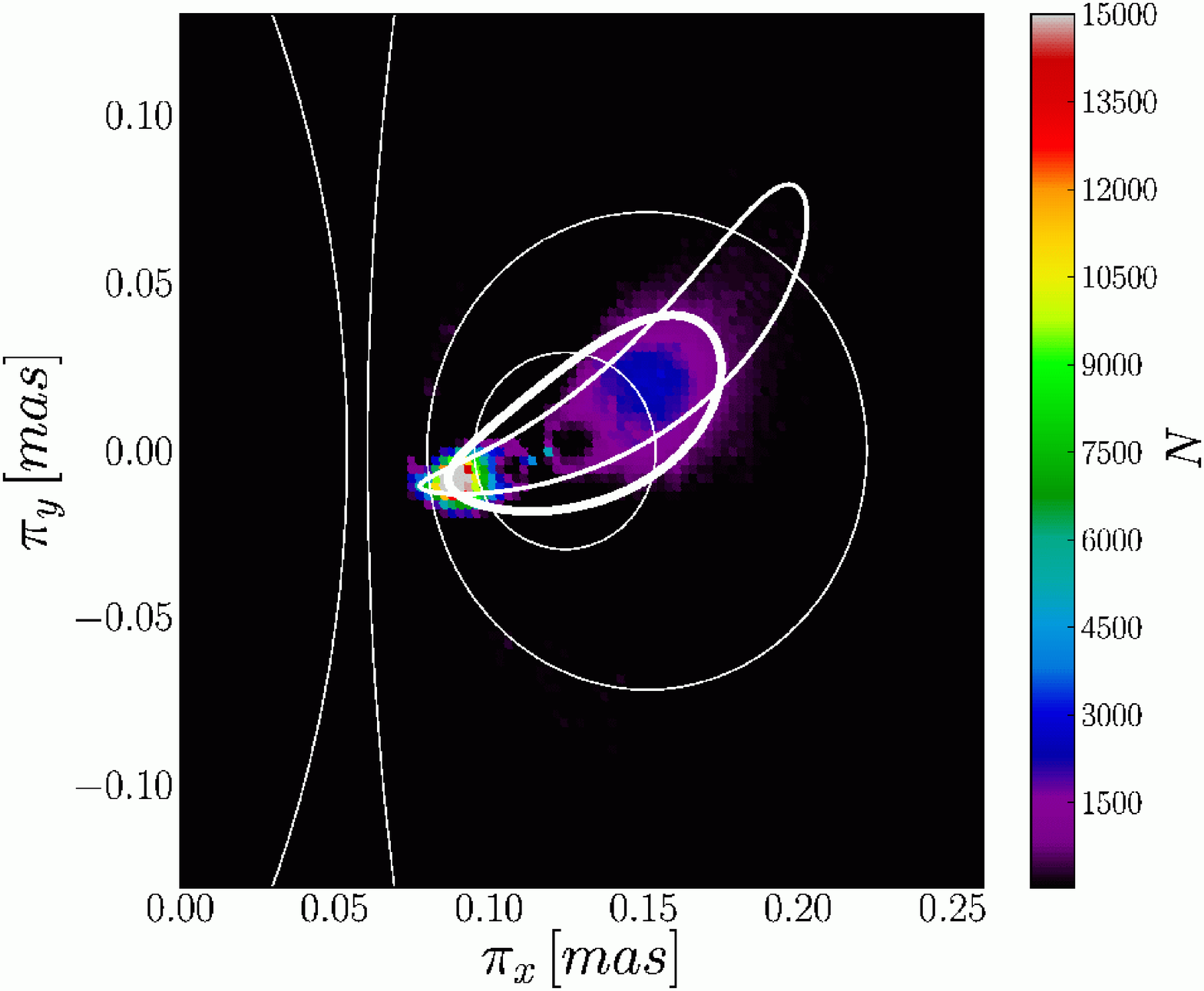}
\end{center}
\caption{ Mapping the RC-all sample in the parallax observable space. 
Top: We plot the whole sample. Note that the colour bar is in log scale. 
Bottom: We plot only the non-axisymmetric component. Note that the colour
bar is linear counts, N, per area of size $2.6\mu as^2$. In both plots we over-plot the structure given by 
the Bulge ellipsoid (thick white line), the 
Long ellipsoid (thin white line) and circles of radius $2$, $4$, $8$ and 
$10\,kpc$, respectively from inside out, to help the reader guide in the 
figure. Note the different colour scale in the plots.}
\label{fig:pigrid}
\end{figure}

In Figure~\ref{fig:pigrid} we map the stellar surface density of the 
RC-all sample in the $(\pi_x,\pi_y)$ plane. We choose a square grid from 
$0.-0.26$mas in order to include all the bar structure as seen in 
Fig.~\ref{fig:elipse} and of $2.6$ $\mu$as of width, which is of the order
of Gaia's resolution. As in Fig.~\ref{fig:Nvol_samples}, we discard stars 
with $R>10$kpc. Due to the relation between distance and parallax, the 
overdensity in the parallax plane does not directly correspond to the 
overdensity in the configuration space. In Fig.~\ref{fig:pigrid}, we can
observe how the particles organize in the expected structures of the bottom
panel of Fig.~\ref{fig:elipse}. The far side of the bar and the far side 
of the Galaxy concentrates at smaller parallaxes, while the near side of 
the bar is more diffuse at higher parallaxes. Nevertheless, we still 
observe the tilted deformation corresponding to the near side of the bar at 
$\pi_x\in (0.15,0.2)$mas.

 The following procedure enables us to subtract the
axisymmetric component in the $(\pi_x,\pi_y)$ plane: 1) We ``axisymmetrize'' the original distribution of test particles in the 
configuration space. To obtain this ``axisymmetrized'' distribution, we convert 
the $(x,y)$ coordinates of each particle to $(R,\theta)$. We keep the $R$, but 
we substitute $\theta$ by a random number between $0$ and $2\pi$. And we 
convert back to $(x,y)$ coordinates. 2) We transform the ``axisymmetrized'' 
distribution to the parallax plane and we count the particles in each cell 
($N_a$). 3) We subtract from the original distribution of test particles in the 
parallax plane ($N_o$) the one from the ``axisymmetrized'' distribution ($N_a$).
When using this strategy to the RC-all sample, we can see in 
Fig.~\ref{fig:pigrid} (bottom panel) that the non-axisymmetric 
component, i.e. the Galactic bar, is well enhanced in the $(\pi_x,\pi_y)$.  

\begin{figure*}
\begin{center}
\includegraphics[scale=0.23]{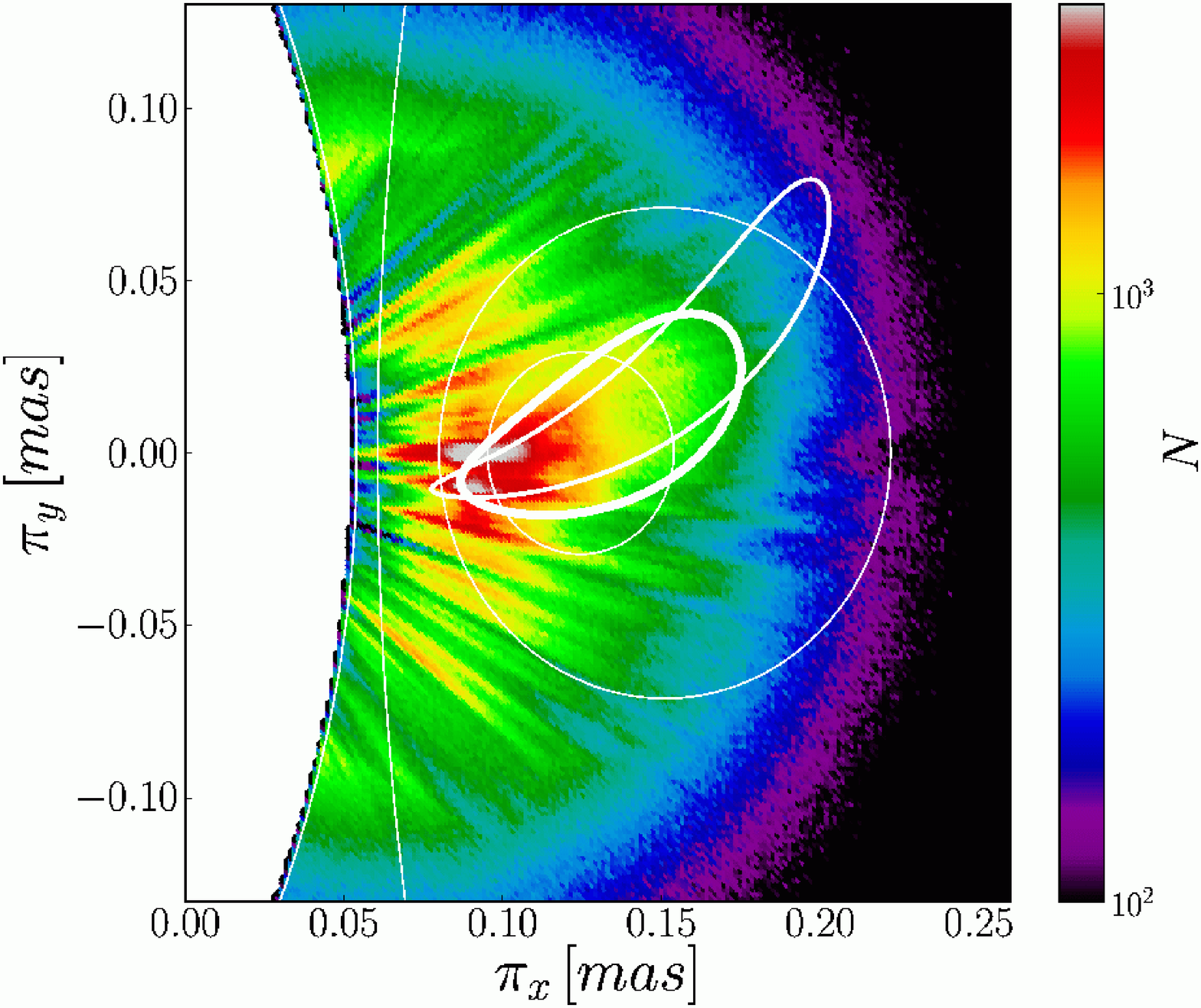}
\includegraphics[scale=0.23]{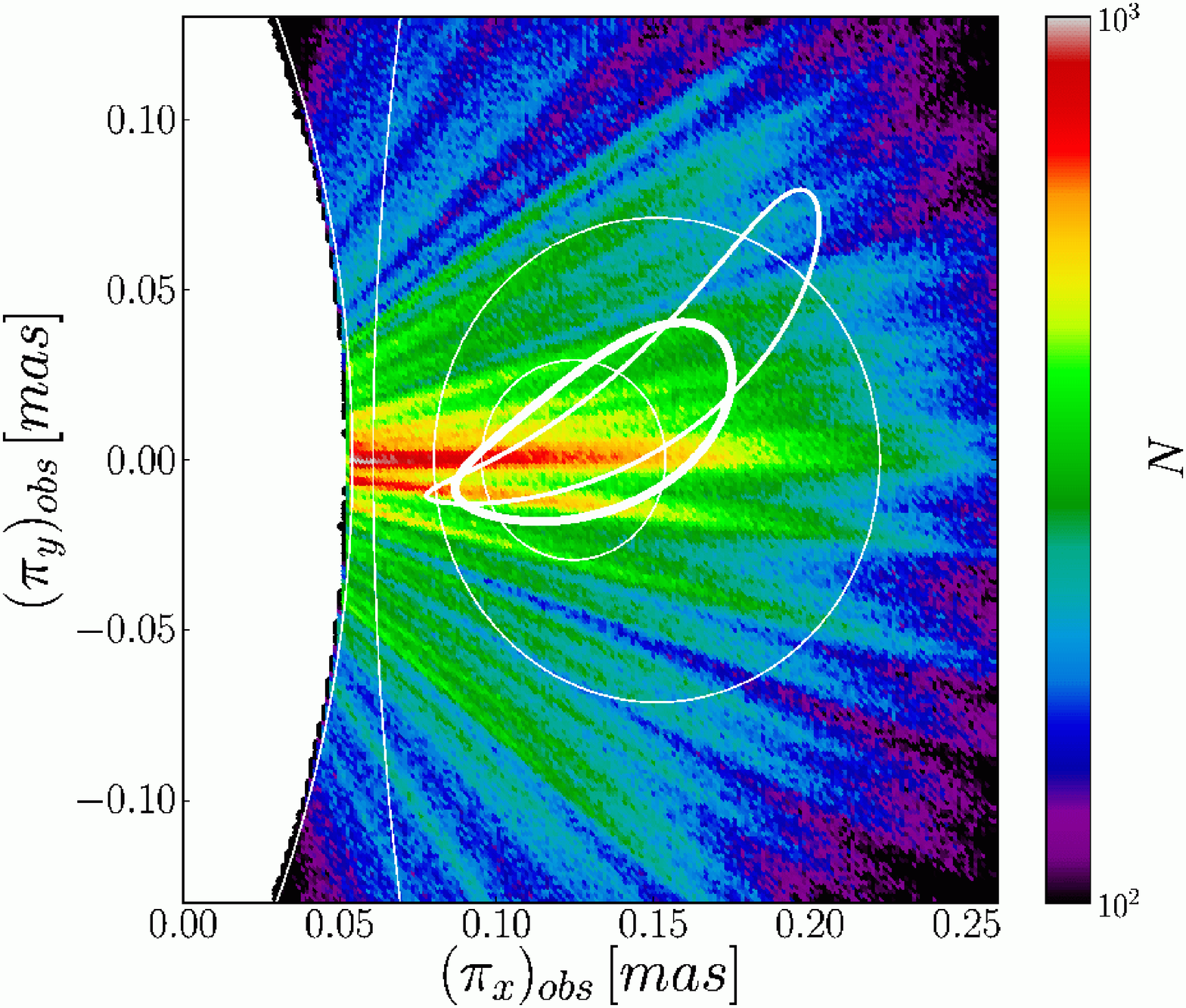}\\
\includegraphics[scale=0.23]{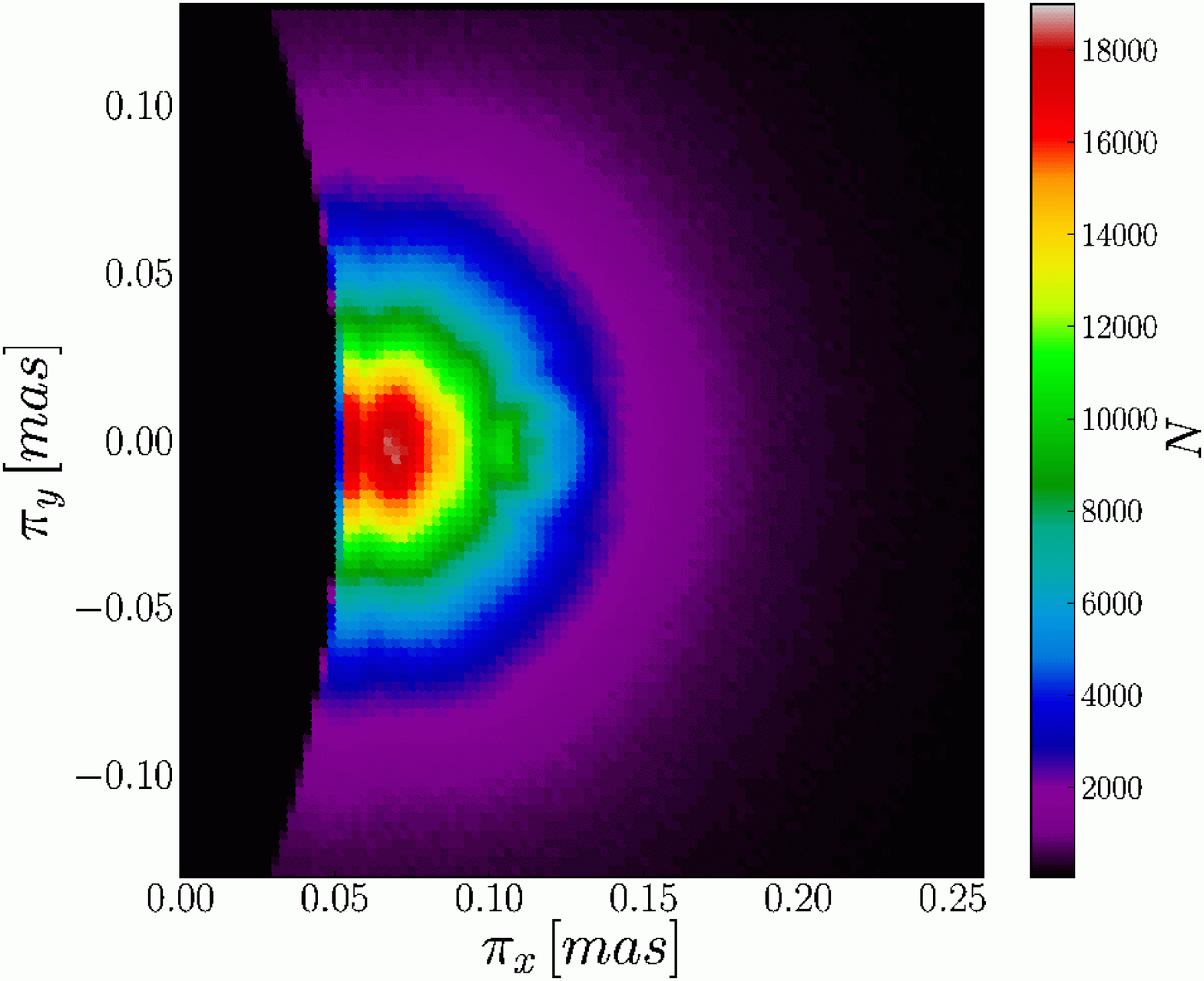}
\includegraphics[scale=0.23]{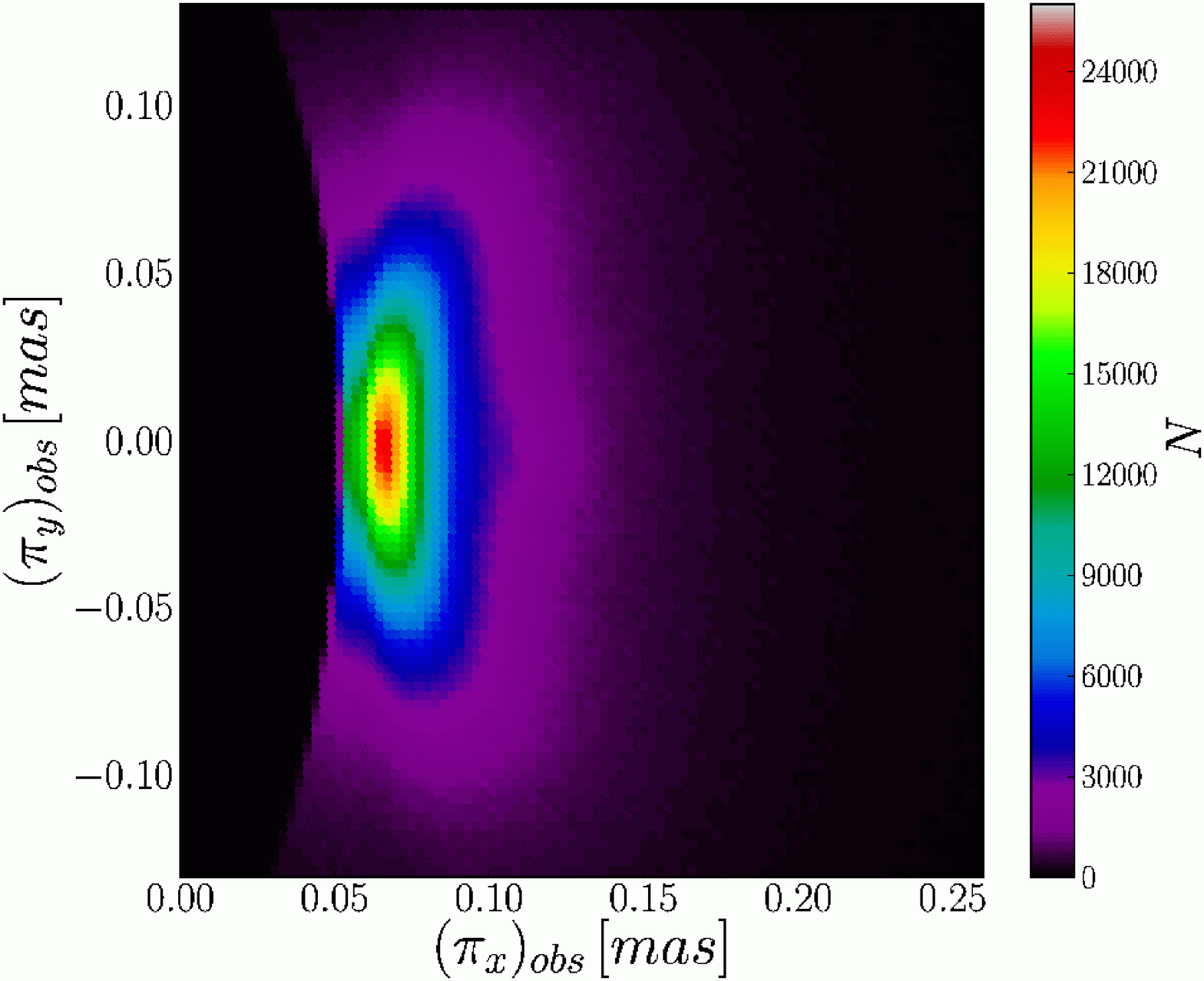}\\
\includegraphics[scale=0.23]{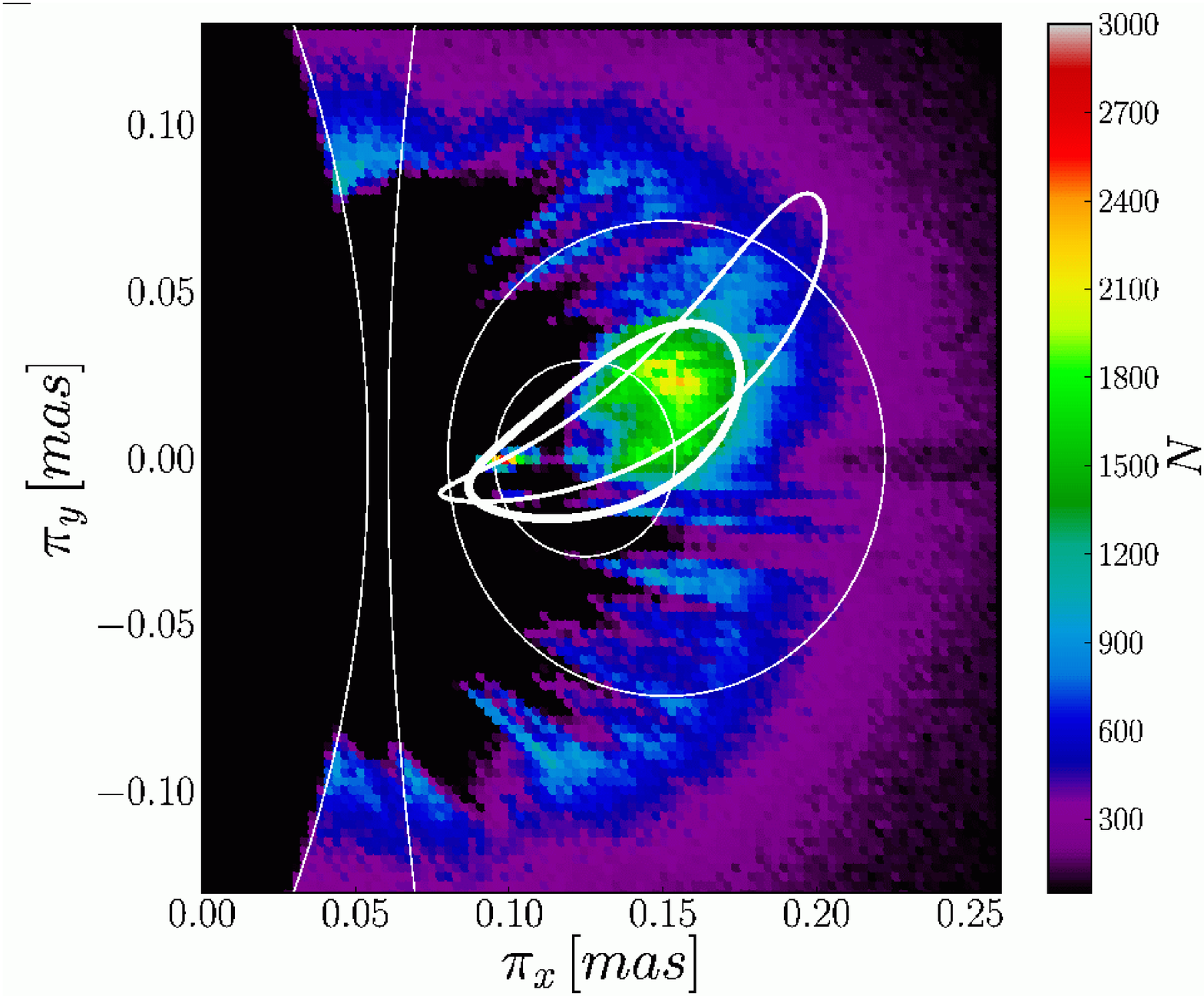}
\includegraphics[scale=0.23]{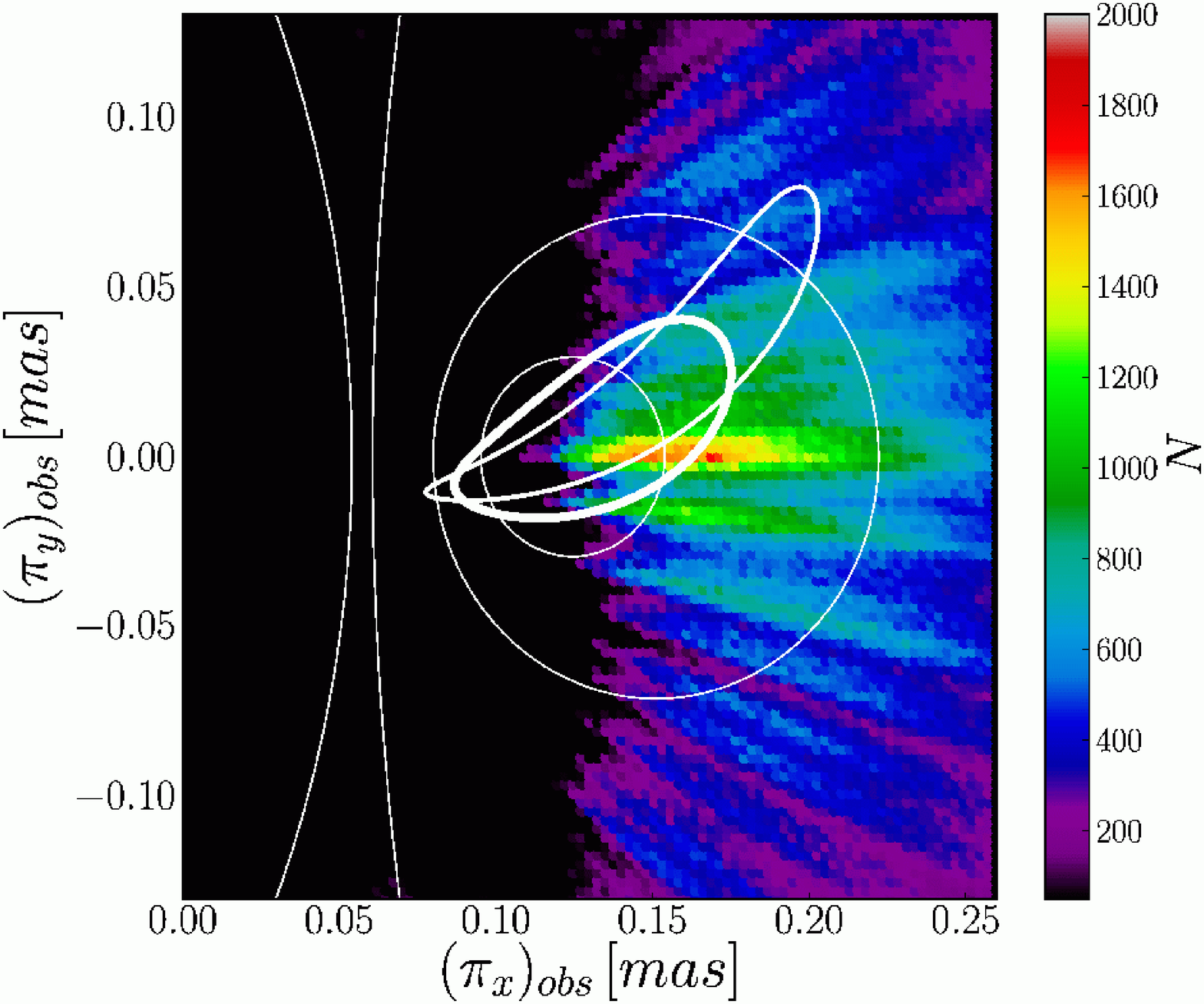}
\end{center}
\caption{ Mapping the RC-G20 sample in the parallax observable space. 
Top panels: RC-G20 sample, i.e. parallaxes without errors (left) and RC-G20-O 
sample, i.e. parallaxes with Gaia errors (right). Middle panels: The 
``axisymmetrized'' distribution for the RC-G20 (left) and
the RC-G20-O (right) samples.
Bottom panels: the non-axisymmetric components of the samples, as the
results of subtracting the ``axisymmetrized'' distributions (second row) from 
the original distributions (top row). 
Note that the colour bar is in log scale (in the top panels) and in 
linear scale (in the middle and bottom panels) and they are counts, N, 
per area of size $2.6\mu as^2$. Note, also, that they do not have the same 
scale. We overplot in the top and bottom panels the structure given by the 
Boxy/bulge ellipsoid (thick white line), the Long ellipsoid 
(thin white line) and circles of radius $2$, $4$, $8$ and $10$kpc, 
respectively from inside out, to help the reader guide in the figure.
We show only data within $R<10$kpc in all panels.}
\label{fig:RC-G20pigrid}
\end{figure*}

Figure~\ref{fig:RC-G20pigrid} top shows the parallax plane for the RC-G20 
sample, i.e. using real parallaxes with no errors (left panels), and the 
RC-G20-O sample, i.e. parallaxes affected by Gaia errors (right panels).
The middle panels show the ``axisymmetrized'' distribution and the
bottom panels show the result of subtracting the ``axisymmetrized'' 
distribution (middle panel) from the original one (top panel), that is, the
bottom panels show the non-axisymmetric components in the parallax plane.
A dominant feature is worth mentioning here, that is, the spikes in surface
density at constant galactic longitude $l$ correspond to galactic longitudes 
with high visual absorption in the Drimmel extinction map \citep{dri03}. 
Secondly, and more important, the effect
of introducing the Gaia errors is clearly seen when we compare both columns. 
When we introduce the Gaia astrometric error in parallax each star expands 
symmetrically in the direction of constant longitude. This is not the case in 
configuration space, because of the fact that the error distribution in 
parallax translates into a skewed error distribution in distance (see 
Fig.~\ref{fig:error_SGC}). This effect is clearly observed in 
Fig.~\ref{fig:RC-G20pigrid} right panels. First, and as an example, 
the overdensity seen in the top left panel, close to the $l=0\gr$ galactic 
longitude (gray ellipse) is spread all over the $\pi_x$ axis when the Gaia 
errors are introduced. Second, it introduces an artificial effect when we 
want to 
subtract the axisymmetric component because the stars affected by error 
in parallax are not symmetrically shuffled in galactocentric radius, which translates in a 
distorted ``axisymetrized'' distribution (see middle right panel of 
Fig.~\ref{fig:RC-G20pigrid}). Therefore, the visual detection 
of the Galactic bar when Gaia parallax errors are considered is not possible. The relative error in parallax 
in the near side of the bar is too big -- up to 40-70\%-- that the 
signature of the bar is lost. 
Once stated the difficulty to recover the Galactic bar characteristics in 
the RC-G20-O sample,  even after subtracting the axisymmetric component,
we consider that more complex methods are required,
such as image reconstruction techniques. As a first step in this direction, 
in the next Sect.~\ref{sec:bar} we take the RC-G20-IR sample, with IR additional
information, so more accurate distances.

 Finally we have to take into account that the method applied above 
to model the axixymmetric compoment could cause non-negligible over- and 
under-substraction when applied to the RC-G20 and RC-G20-O samples. Due to 
the observational constraints, namely the cut in apparent magnitude (so 
distance) and the important dust extinction in the galactic plane, these 
samples present an important irregular surface density around a given 
galactocentric ring. Large foreground extinction along a given 
line-of-sight direction would reduce the total number of observed stars, thus, 
the substraction of a mean axisymmetric component will be overestimated in 
this position. On the contrary, areas well sampled by Gaia will have an 
under-substraction of the axisymmetric component as other parts of the 
ring are not well sampled, so the mean axisymmetric component is lower 
than the real one. Despite this drawback, this does not prevent from
detecting the Galactic bar when using IR photometric distances.

\subsection{The angular orientation of the Galactic bar using IR data}
\label{sec:bar}
\begin{figure}
\begin{center}
\includegraphics[scale=0.3]{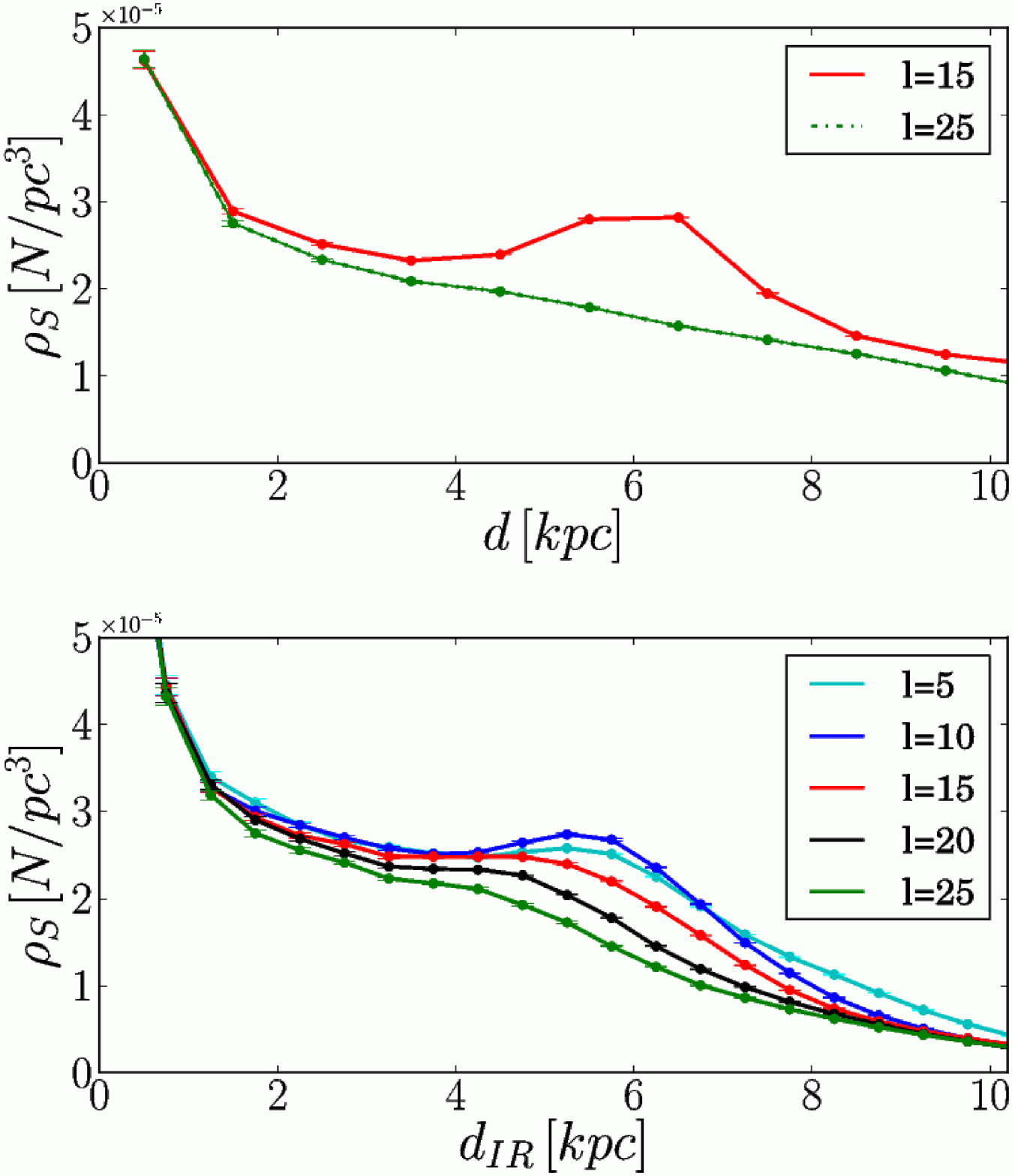}
\end{center}
\caption{Volume density as a function of the heliocentric distance at 
different lines-of-sight (with a solid angle of $4\gr$). Top panel: 
lines-of-sight ($l=15$ in red and $l=25$ in green) for the RC-all sample.
Bottom panel: five lines-of-sight ($l=5$ in cyan, $l=10$ in blue, $l=15$ 
in red, $l=20$ in black and $l=25$ in green) for the RC-G20-IR
sample.}
\label{fig:Nvol_dist}
\end{figure}

 Working in the configuration space and using IR distances, we are 
interested in not only detecting the Galactic bar but also
determining its angular orientation. In Fig.~\ref{fig:Nvol_dist} we show
the spheric volume density, $\rho_S$, as a function of the heliocentric 
distance for the RC-all (top panel) and the RC-G20-IR (bottom panel) samples 
for several lines-of-sight. We select the stars within $l\pm 2\gr$ and in 
distance bins of $500$pc, we then divide the number of stars by the difference 
of volume of the spherical wedges of two consecutive distance bins. This is, 
$\rho=N_s/V$, where $N_s$ is the number of stars, 
\begin{equation}
V=\frac{\alpha}{360}\frac{4\pi}{3}\left(d_{i+1}^3-d_i^3\right),
\end{equation}
$\alpha=4\gr$ is the solid angle of the wedge and $d_i$ and $d_{i+1}$ are
distances of two consecutive bins in parsecs. Using the RC-all sample, we see 
that at $l=15\gr$, we can clearly see the bar overdensity at about $6$kpc from 
the Solar position, while at $l=25\gr$, there is no overdensity. This is also 
determined using the RC-G20-IR sample with IR errors. Note that the bar 
overdensity is detected from $l=15\gr$ towards inner lines-of-sight. Even 
though these plots are useful to detect the bar overdensity, they are not 
precise or robust enough to determine the angular orientation of the bar.

\begin{figure}
\begin{center}
\includegraphics[scale=0.23]{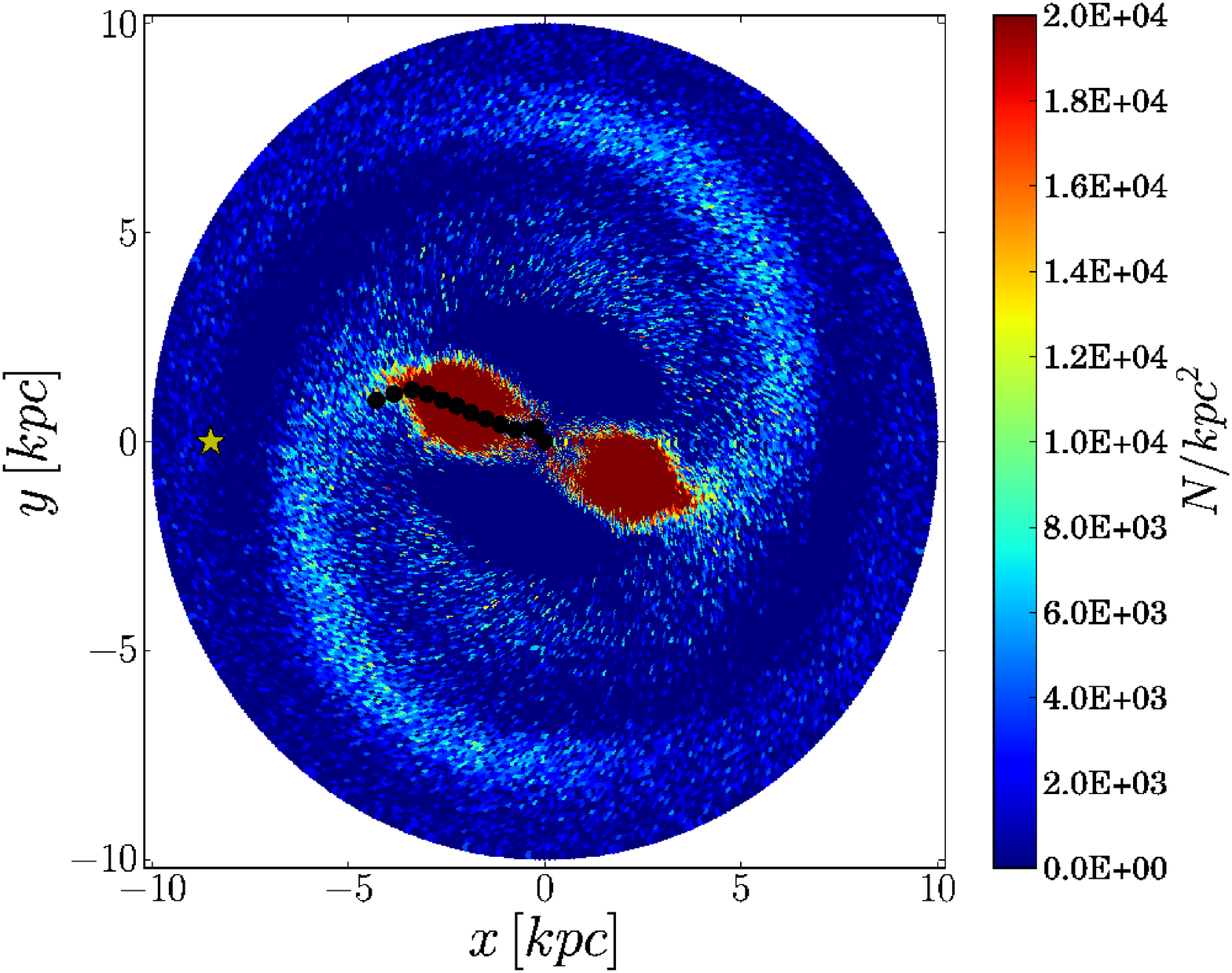}\\
\includegraphics[scale=0.23]{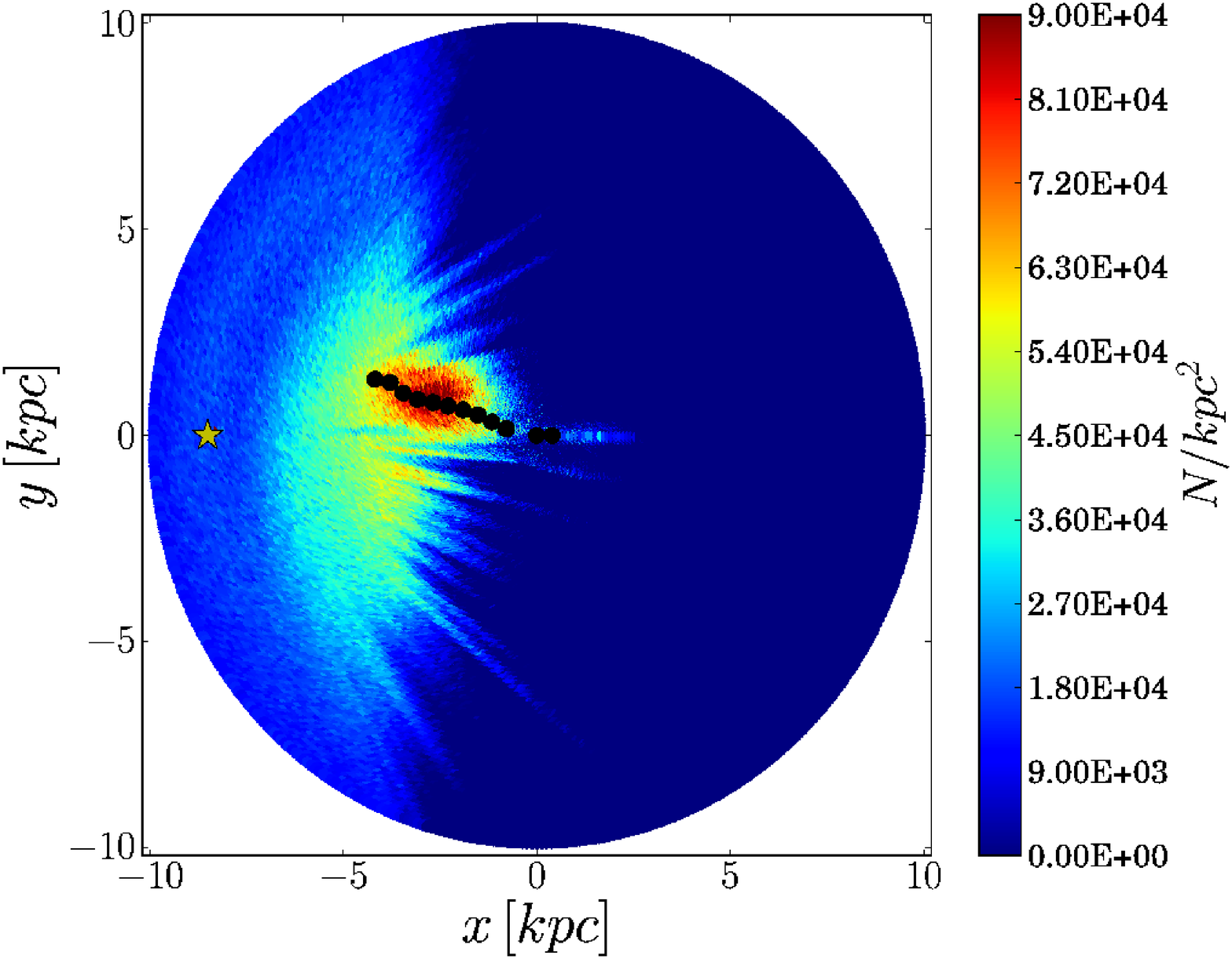}
\end{center}
\caption{Determination of the bar azimuthal angle. Top panel: RC-all sample
with the axisymmetric component subtracted. Bottom panel: RC-G20-IR sample
also with the axisymmetric component subtracted. The back dots show the
result of the Gaussian fit in the determination of the bar azimuth. 
 Note that the scale of the colour bar is different in both panels.}
\label{fig:bar}
\end{figure}

We then develop a method to determine the azimuthal angle of the bar 
overdensity in our Gaia samples following two steps. First, we subtract 
the axisymmetric
component to enhance the bar structure, and, second, we perform a Gaussian
fit to the subtracted stellar overdensity to locate the angular position
of the maximum stellar density. The procedure is as follows:
\begin{enumerate}
\item Subtraction of the axisymmetric component. We make galactocentric 
radial bins of $100$pc and we compute the mean surface density inside each
ring. We then subtract the mean from the initial surface density map. 
From that we obtain an image with the non-axisymmetric components clearly
enhanced (see top 
panel of Fig.~\ref{fig:bar} for the RC-all sample).
\item Gaussian fit to the surface stellar overdenity. We make 
galactocentric polar bins of $400$pc in radius and $0.72\gr$ of azimuthal width.
For each radial ring, we fit a Gaussian function to the data using least 
squares. This allows us to derive the galactocentric azimuth at which
the function has an absolute maximum (the mean of the fitted Gaussian). The 
 error bar assigned to the azimuth of the maximum density is the $1\sigma$ 
error derived from the least square of the Gaussian fit.
\end{enumerate}

\begin{figure}
\begin{center}
\includegraphics[scale=0.27]{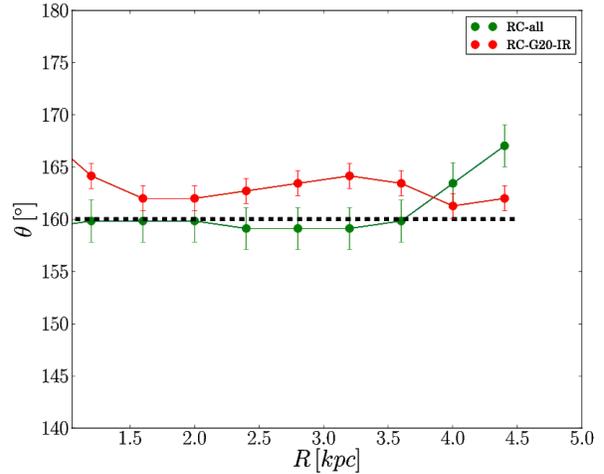}
\end{center}
\caption{Determination of the bar azimuthal angle. We show in polar
coordinates the result of the Gaussian fit to the RC-all sample (green curve) 
and RC-G20-IR sample (red curve). The horizontal thick black line corresponds
to the imposed azimuthal angle in the model. The azimuth is defined positive 
counter-clockwise and from the x-positive axis. The bar, therefore, is 
located at $160\gr$.}
\label{fig:curve}
\end{figure}

This procedure has been applied to the RC-all and RC-G20-IR samples
(Figs.~\ref{fig:bar} and \ref{fig:curve}). The RC-all sample allows us to 
check the performance of the method. In the top panel of Fig.~\ref{fig:bar}, 
we show the surface density resulting after the subtraction of the 
axisymmetric part and the black dots are the result of the Gaussian fit.
We perform the Gaussian fit to the stars in the first heliocentric 
quadrant ($0\le l \le 90\gr$), which is where the near side of the Galactic 
bar is located. As expected, we clearly detect the Galactic bar, but we also 
highlight the inner ring and the response spiral arms due to the presence of 
the invariant manifolds \citep{rom07}. Note that the black dots follow 
the bar semi-major axis up to $R\sim 4$kpc, which corresponds to a radius
intermediate between the one of the boxy/bulge bar and the one of the Long bar. 
This is more clearly shown in polar coordinates in Fig.~\ref{fig:curve}. 
The azimuthal angle is defined positively in the counter-clockwise 
direction from the x-positive axis. Thus, the points line up along the 
$160\gr$ constant azimuthal angle. The discrepancy at the end of 
the bar is clearly observed after $R>3.5$kpc associated to the overdensity
of the inner ring and the spiral arms. This discrepancy would indicate 
approximately the length of the bar.

We then apply the same procedure to the RC-G20-IR sample (see bottom panel
of Fig.~\ref{fig:bar}). First, note that again, the extinction blurs the
contribution of the non-axisymmetric components to the surface density. 
However, we still can observe the near side of the Galactic bar.  
If we examine this in polar coordinates (red curve in Fig.~\ref{fig:curve}), 
we note the good recovery of the bar azimuthal angle. We observe a small
bias of $\sim 3-5\gr$, deviated with respect to the nominal value in
about $2-3\sigma$. Several factors account for this bias. 
First, the RC-G20-IR sample has the effects of the extinction model. Given a 
certain line-of-sight, the number of stars observed will decrease with the
heliocentric distance. This fact can translate into a change of the bar maximum 
observed density towards higher values of $\theta$. 
Second, the fact that the RC-G20-IR sample is magnitude-limited makes that 
intrinsically brighter stars are over-represented, this is the Malmquist bias. 
This can also lead to biased values of the azimuthal angle, which are not 
trivial to correct \citep{are02}. 
Third, there is also a geometric bias due to the fact that even having a 
symmetric error in photometric distance along the line of sight, it translates
into a non-symmetric $\Delta \theta$ from the Galactic Centre. Only stars in a 
line-of-sight perpendicular to the semi-major axis of the bar will not suffer 
from this bias.
In any case, even taking into account the possible biases, the
Gaussian fit method recovers well the azimuthal angle of the Galactic bar.

\section{Conclusions}
\label{sec:conc}
In this first paper of a series, we present two Red Clump Gaia mock 
catalogues. They are obtained as the integration of a set of test
particles using a 3D barred galaxy model. We set the particles to 
mimic the characteristics of the disc RC giant stars and we use the 
\citet{dri03} extinction model with rescaling factors to obtain the apparent 
Gaia magnitudes $G$ and $G_{RVS}$ of the particles. Adding the Gaia error 
model available before commissioning for astrometry, photometry and 
spectroscopy, we obtain the corresponding observed catalogues, RC-G20-O and 
RC-RVS-O. The first catalogue, RC-G20-O, contains all particles observed
by Gaia, i.e. we only apply a cut in magnitude ($G<20$), which reduces the 
number of particles by a factor of 2 with respect to the particles initially 
generated. Thus, Gaia will observe about $2.6\times 10^7$ RC disc stars. 
The second catalogue, RC-RVS-O, includes a second cut by requiring that the 
errors in radial velocity are small, $\sigma_{V_r}\le 10$\kms in the nominal 
performances. The number of particles of this sample is $8.5\times 10^6$. 
This second mock catalogue contains all the 6 coordinates of phase space 
with good precision, since at an heliocentric distance of $4$kpc, the mean 
errors in tangential and radial velocity are, respectively, $\sim 20$\kms 
close to the Galactic plane and $\sim 10$\kms above $|z|>300$pc and 
$\sim 6$\kms close to the Galactic plane and $\sim 4$\kms, again above 
$|z|>300$pc.

We present a first attempt to describe the Galactic bar 
in the Gaia observable space, this is the cartesian projection of the 
$(\pi,l)$ coordinates $(\pi_x,\pi_y)=(\pi\cos l,\pi\sin l)$. We first analyse 
how the Galactic bar is projected in this space and then we apply this 
transformation to the RC-G20-O sample. We conclude that the Gaia relative 
errors in parallax and the high interstellar extinction in the inner parts 
of the Galactic disc prevent us to model the bar overdensity. This space 
is the natural space to directly see the effect of the extinction and we 
confirm the difficulties of Gaia, which works in the optical, to trace the 
bar structure without a robust statistical treatment. 

This result points towards the need to combine Gaia and IR data to undertake 
the study of the internal disc structures. Since RC stars photometric distance 
has a better precision using IR surveys, we convolve the Gaia RC-G20 catalogue 
with a constant relative error in distance (we assume $10\%$ as in APOGEE 
\citep{bov14}), obtaining the RC-G20-IR sample. In this case, the quality of 
the distances is good enough to clearly detect the bar overdensity, as seen 
in Figs.~\ref{fig:bar} and \ref{fig:curve}, and to allow an estimation of 
the bar angular orientation with an accuracy of $\sim 5\gr$. In this work we
quantify how the combination of surveys opens new avenues for the 
studies of the Galactic disc, in particular of the signatures of the 
Galactic bar. 

 The tool presented here can be adapted to any type of star. The study of
the Galactic bar considering only the RC stars population has proven to be 
complex. However, Gaia will detect intrinsically brighter and redder
late type giant stars, such as $M$ giant stars, with better accuracy than 
the RC stars at the same position. Assuming $M_V=-0.4$ and $(V-I_c=1.78)$ as in \citet{hun14},
it gives for a star at about $4$kpc from the Sun, a $\sigma_{\pi}\sim20\mu$as,
and a relative error in parallax of about $5\%$, using the before commissioning 
error model. This good precision could help improving and complementing this 
work in detecting the Galactic bar overdensity. 

The signature of the bar is present not only in density but it also shows 
imprints on the kinematic space, i.e. forming moving groups
\citep[e.g.][]{deh00,fux01,gar10,min10,ant11}. Therefore, with Gaia, we are
not limited to the study of the bar overdensity, but we can use all the
6D phase space. We are currently working on the analysis of the moments of 
the velocity distribution function in the Gaia sphere, about $4-5$kpc from 
the Sun, to try to obtain information on the potential of the Galaxy 
(Romero-G\'omez et al, in preparation).
 
In this work, we have made the first attempts in evaluating the future 
Gaia data of RC stars and the capabilities of simple tools to
recover the characteristics of the Galactic bar overdensity. Future work 
has to be done in developing appropriate tools, such as the application of
statistically robust image reconstruction techniques in the space of the Gaia
observables, where the contribution of the error can be easily accounted for.

\section*{Acknowledgements}
This work was supported by the MINECO (Spanish Ministry of Economy) - 
FEDER through grant AYA2012-39551-C02-01 and ESP2013-48318-C2-1-R. 
 We thank the anonymous referee for his/her useful comments that helped
improving the manuscript.
HA acknowledges the support of the Gaia Research for European Astronomy 
Training (GREAT-ITN) network funding from the European Union Seventh 
Framework Programme, under grant agreement $n\degr$264895.
LA thanks the Gaia community at the University of Barcelona for their 
hospitality and the Spanish Ministry of Education for the sabbatical grant 
SAB2010-0120.

\appendix
\section{The generation of the Initial Conditions}
\label{sec:appA}
The initial conditions follow the density distribution of the \citet{miy75}
disc, because this is the density distribution chosen in 
\citet{all91} to characterise the disc of the Galaxy. The fact that the 
parameters of the disc defining the initial conditions are the same as the
disc of the axisymmetric component used in the integration facilitates the
relaxation of the particles. Particles generated using a mass distribution
similar to the mass distribution imposed in the integration reach statistical
equilibrium faster. The initial conditions are generated using the Hernquist 
method \citep{Hernquist.93}. Here we summarize the steps:

\subsection{Radial distribution}
\label{sec:posR}
We first compute the normalized cumulative distribution function
\begin{equation}
\S^*(R)=\frac{1}{\Sigma_0}\int_0^R\Sigma^*(R^{\prime})dR^{\prime},
\end{equation}
where $\Sigma^*(R)=R\int_{-\infty}^{+\infty} \rho(R,z)dz$ is the probability 
distribution function in cylindrical coordinates for the radial component, 
$\Sigma_0$ is the normalization constant, taken as 
$\Sigma_0=\int_0^{+\infty}\Sigma^*(R^{\prime})dR^{\prime}$, and $\rho(R,z)$ is
the density of the Miyamoto-Nagai disc.

Once we have the normalized cumulative distribution function for a discrete
set of radii, we use its inverse to generate the radial positions. 
Once we have the galactocentric distance of the star, R, we can obtain the 
cartesian coordinates $(x,y)$ by generating a random number in $[0,2\pi)$ and 
applying the polar coordinate transformation.

\subsection{Vertical distribution}
\label{sec:posz}
The generation of the vertical coordinate $z$ is performed in a similar way
as in \S~\ref{sec:posR}. The distribution function in z is derived from the 
integration of the vertical Jeans equation, for the particular case when 
$\sigma_z^2=\sigma_z^2(R)$, i.e. the vertical velocity dispersion only 
depends on $R$. For a given radius, $R$, the probability 
distribution function for the coordinate $z$ is
\begin{equation}
\nu(R,z)=e^{\Delta\Phi(z)/\sigma_z^2(R)},
\end{equation}
where $\Delta\Phi(z)=\Phi(R,0)-\Phi(R,z)$ and $\Phi(R,z)$ is the Miyamoto-
Nagai potential, and $\sigma_z^2(R)$ is the vertical velocity dispersion 
taken as, $\sigma_z^2(R)=\pi Gz_0\Sigma(R)$,
where $z_0$ is the scale-length in z, considered constant here (as a first
approximation of the solution) and $\Sigma(R)$ is the Miyamoto-Nagai
surface density, computed numerically.

To randomly obtain the coordinate $z$, we use the Von 
Neumann Rejection Technique using this probability distribution function 
\citep{pre92}.

\subsection{Generating the velocities}
\label{sec:velocities}
Now we need to generate the velocities associated to the positions generated
above. First we define the radial, tangential and vertical dispersions.

As for the radial velocity dispersion, it consists of setting the square 
of the velocity dispersion proportional to the surface density and 
normalizing to a given value. We fix the value of $\sigma_U$ at a the
Solar radius. This gives:
\begin{equation}
\sigma_U(R)=\sigma_U(R_{\odot})\left(\frac{\Sigma(R)}{\Sigma(R_{\odot})}\right)^{1/2}.
\end{equation}
Note that we are assuming that the radial scale-length is double that of the
density distribution.

The tangential velocity dispersion, $\sigma_V$, is determined by assuming 
the epicyclic approximation, that is:
\begin{equation}
\frac{\sigma_V^2(R)}{\sigma_U^2(R)}=\frac{\kappa^2}{4\Omega^2},
\end{equation}
where $\kappa$ is the epicyclic frequency, and $\Omega$ is the angular 
frequency, computed from the rotation curve of the \citet{all91} potential.

Finally, the vertical velocity dispersion, $\sigma_W$, is also set to
proportional to the square root of the surface density. This expression
comes from the Jeans and Poisson equations, together with the Eddington
approximation (no coupling between $R$ and $z$) and isothermality in the
vertical direction, that is $\sigma_W$ is independent of $z$.
\begin{equation}
\sigma_W(R)=\left(\pi G z_0 \Sigma(R)\right)^{1/2},
\end{equation}
where $z_0$ is the scale-height of the disc, here considered constant, and
$\Sigma(R)$ is the surface density of the Miyamoto-Nagai disk. 

We now generate residual velocity components of each particle at position 
$(R,z)$ with respect to the Regional Standard of Rest, $(U,V,W)$, using a 
Gaussian with the respective velocity
dispersions. There is though one final step to consider. We need to add the 
circular velocity, according the \citet{all91} potential, and subtract the 
asymmetric drift to the tangential
component. The asymmetric drift is approximated as \citep{bin08}
\begin{equation}
V_a=\frac{1}{2V_c}\left[\left(\frac{\kappa^2}{4\Omega^2}-1-\frac{R}{\Sigma}\frac{\partial\Sigma^{\prime}}{\partial R}\right)\sigma_U^2-R\frac{\partial \sigma_U^2}{\partial R}\right],
\end{equation}
where $\Sigma^{\prime}$ is the Miyamoto-Nagai density cut at $z=0$ and $V_c$ is
the circular velocity both at the given radius $R$.

\bibliography{GaiaRC}

\label{lastpage}

\end{document}